\documentclass[11pt]{article}
\usepackage{amssymb,amsmath}
\usepackage{bm} 
\usepackage{booktabs} 
\usepackage{array}
\usepackage{latexsym}
\usepackage{graphicx}
\usepackage{color}
\usepackage{datetime}
\usepackage[nosort]{cite}
\usepackage{verbatim}
\usepackage{enumerate}
\usepackage{chngpage} 
\usepackage{mathrsfs}
\usepackage{euscript}
\usepackage{psfrag}
\usepackage{subfig}
\usepackage{anyfontsize}
\usepackage{braket}

\usepackage{blindtext}%
\usepackage[theorems,breakable]{tcolorbox}%

\usepackage{multirow,graphicx}
\usepackage{makecell}
\usepackage{float}
\usepackage{tikz}
\usetikzlibrary{decorations.text}
\usepackage{enumitem}
\setlist{itemsep=0pt}
\usetikzlibrary{math}
\usepackage{setspace}

\usepackage{mciteplus}

\usepackage[colorlinks=true,      linkcolor=blue,      urlcolor=blue,      
            filecolor=blue,      citecolor=blue,       pdfstartview=FitH,     
						pdfpagemode=UseNone,      bookmarksopen=true]{hyperref}  
\usepackage[all]{hypcap}     

\usepackage{tikz}
\usepackage{pgfplots}
\usetikzlibrary{decorations.text,intersections, pgfplots.fillbetween}
\usetikzlibrary{math}
 \usetikzlibrary{3d,shapes.geometric,shadows.blur}
\usepackage{tikz-3dplot}

\usepackage{blindtext}%
\usepackage[theorems,breakable]{tcolorbox}%

\newtcbtheorem{mytheo}{}%
{colback=gray!5,colframe=gray!35!black,fonttitle=\bfseries}{th}

\newtcbtheorem{lem}{}{%
        colback=green!5,%
        colframe=green!35!black,%
        fonttitle=\bfseries,title %
    }{lem}

\definecolor{amaranthred}{rgb}{0.83,0.13,0.18}
\definecolor{amazon}{rgb}{0.23,0.48,0.34}
\definecolor{bdazzledblue}{rgb}{0.18,0.35,0.58}
\definecolor{absolutezero}{rgb}{0.0,0.28,0.73}
\definecolor{bitterlemon}{rgb}{0.79,0.88,0.05}
\definecolor{byzantine}{rgb}{0.74,0.2,0.64}
\definecolor{turquoise}{rgb}{0.19, 0.84, 0.78}
\definecolor{burgundy}{rgb}{0.5, 0.0, 0.13}
\definecolor{airforceblue}{rgb}{0.36, 0.54, 0.66}
\definecolor{arsenic}{rgb}{0.23, 0.27, 0.29}

\newcommand{\comm}[1]{} 

\def\({\left(}
\def\){\right)}
\def\[{\left[}
\def\]{\right]}

\def\One{{\hbox{ 1\kern-.8mm l}}}

\def\barray{\begin{array}}
\def\earray{\end{array}}
\def\be{\begin{equation}}
\def\ee{\end{equation}}
\def\bea{\begin{eqnarray}}
\def\eea{\end{eqnarray}}
\def\bal{\begin{align}}
\def\eal{\end{align}}
\def\nn{\nonumber}

\def\-{\,-\,}
\def\={\,=\,}
\def\+{\,+\,}
\def\equi{\,\equiv\,}

\numberwithin{equation}{section} 

\definecolor{cardinal}{rgb}{0.6,0,0}
\definecolor{darkgreen}{rgb}{0,0.4,0}
\definecolor{golden}{rgb}{0.92, 0.7, 0}
\definecolor{midnight}{rgb}{0, 0, 0.5}
\definecolor{darkblue}{rgb}{0, 0, 0.7}
\definecolor{purple}{rgb}{0.5, 0, 0.5}

\definecolor{amaranthred}{rgb}{0.83,0.13,0.18}
\definecolor{amazon}{rgb}{0.23,0.48,0.34}
\definecolor{bdazzledblue}{rgb}{0.18,0.35,0.58}
\definecolor{absolutezero}{rgb}{0.0,0.28,0.73}
\definecolor{bitterlemon}{rgb}{0.79,0.88,0.05}
\definecolor{byzantine}{rgb}{0.74,0.2,0.64}
\definecolor{turquoise}{rgb}{0.19, 0.84, 0.78}
\definecolor{burgundy}{rgb}{0.5, 0.0, 0.13}

\def\IR{\mathbb{R}}

\def\cA{{\cal A}}
\def\cB{{\cal B}}

\def\cE{{\cal E}}

\def\cG{{\cal G}}
\def\cH{{\cal H}}
\def\cJ{{\cal J}}
\def\cI{{\cal I}}

\def\cM{{\cal M}}
\def\cN{{\cal N}}
\def\cP{{\cal P}}
\def\cO{{\cal O}}

\def\cS{{\cal S}}

\def\cV{{\cal V}}
\def\cW{{\cal W}}
\def\cX{{\cal X}}
\def\cY{{\cal Y}}
\def\cZ{{\cal Z}}

\usetikzlibrary{positioning}

\begin{document}

\begin{titlepage}
\thispagestyle{empty}

\begin{flushright}

\end{flushright}

\begin{center}
\vspace{2cm}
\noindent{\bf \Large Microstates of Non-Extremal Black Holes: }\\
\vspace{0.25cm}
\noindent{\bf \Large A New Hope}\\
\vspace{0.2cm}

\vspace{1cm}

{\bf \normalsize Soumangsu Chakraborty and  Pierre Heidmann}
\vspace{1cm}\\

{\it
Department of Physics and Center for Cosmology and AstroParticle Physics (CCAPP), \\
The Ohio State University, Columbus, OH 43210, USA
}\\[1.5mm]

\vspace{1 cm}

{\footnotesize\upshape\ttfamily heidmann.5@osu.edu,~ soumangsuchakraborty@gmail.com } 

\vspace{1.75cm}
\end{center}

\begin{abstract}

We provide a new roadmap for constructing microstates of non-extremal black holes in supergravity. First, we review the non-linear sigma model of five-dimensional supergravity governing stationary solutions with a U(1) isometry and present the first generalized Ernst formulation of this model.  We then revisit solution-generating techniques associated to the coset model symmetry,  Ernst formalism and inverse scattering method.  While some of these techniques have been extensively used to generate non-extremal black holes and black rings in supergravity, we demonstrate how they can be adapted to construct systematically non-BPS smooth horizonless geometries that have the same mass and charges as non-extremal black holes. To illustrate these methods, we construct novel static solutions of this type, including a non-BPS generalization of the $\frac{1}{2}$-BPS Gibbons-Hawking center, which has served as the fundamental building block of multicenter microstates of BPS black holes.

\end{abstract}

\end{titlepage}

\thispagestyle{empty}

\newpage


\setcounter{tocdepth}{2}
\hrulefill
\vspace{-0.5cm}
\tableofcontents

\hrulefill

\vspace{-0.5cm}
\section{Introduction}

The construction of exact solutions in classical theories of gravity has always been significantly constrained by the challenge of solving Einstein's field equations. These equations are notoriously complex and generally non-separable,  so one typically adopts an ansatz that reduces the equations to an integrable system. Nonetheless, the ability to construct exact solutions remains of fundamental interest, particularly for two key reasons. First, exact solutions serve as vital tests of the uniqueness theorem in classical gravity, helping to determine whether black holes are the only admissible stationary solutions, or if alternative geometries with the same conserved charges can exist. Second, obtaining exact classical solutions has been fundamental for systematically constructing coherent black hole microstates in string theory \cite{Bena:2022rna}, providing crucial insights into the nature of black hole microstructure and its effects on horizon-scale physics. These motivations have driven the development of two particularly successful approaches:
\vspace{-0.1cm}
\begin{itemize}
\item[•] \underline{The sigma-model approach:} In four-dimensional theories of gravity, restricting to stationary solutions (assuming a timelike Killing vector $\partial_t$) reduces the action to a nonlinear sigma model with a characteristic symmetry group: SL(2,$\IR$) for vacuum GR \cite{Maison:1979kx},  SU(2,1) for Einstein-Maxwell theory\cite{Ernst:1967by,Eris:1984tu},  or Sp(4,$\IR$) for Einstein-Maxwell-dilaton-axion theory \cite{Galtsov:1994pd,Galtsov:1997jrl}.  Various solution-generating techniques have been developed based on these integrable structures, including the Ernst formalism \cite{Ernst:1967by}, scattering methods \cite{Belinsky:1971nt,Belinsky:1979mh}, B\"acklund-Ehlers-Harrison transformations \cite{harrison_new_1968,Geroch:1970nt,PhysRevLett.41.1197,Harrison:1980fr,Alekseev:2020bqu}, and the Sibgatullin method \cite{Manko_1993}. These methods have facilitated the derivation of intricate solutions and provided tests of the uniqueness theorem in four dimensions. A striking example of the approach's efficacy is the most general solution in Einstein-Maxwell theory describing a bound state of an arbitrary number of Kerr-Newman black holes with arbitrary masses, charges, and spins \cite{Ruiz:1995uh}.
\item[•] \underline{The supersymmetric approach:} In supergravity theories,  focusing on geometries that preserve some supersymmetry allows to replace Einstein's field equations with integrable first-order equations.  The list of solution-generating techniques is extensive and varies with the chosen supergravity framework and the number of preserved supersymmetries. Unlike in four dimensions,  supersymmetric theories in higher dimensions allow for vast families of smooth, horizonless geometries parameterized by arbitrary functions rather than discrete parameters. Notable examples include the LLM geometries \cite{Lin:2004nb}, Lunin-Mathur solutions \cite{Lunin:2001fv}, and superstrata \cite{Bena:2015bea,Heidmann:2019xrd}, some of which represent coherent microstates of supersymmetric black holes.
\end{itemize}
\vspace{-0.1cm}

Despite their successes, both approaches have limitations. The supersymmetric approach has revealed extensive gravitational degrees of freedom that circumvent the uniqueness theorem, leading to large families of horizonless gravitational solitons. However, it is highly dependent on supersymmetry, and extending these results to more astrophysically relevant, non-supersymmetric setups has been challenging. Conversely, the sigma-model approach in four dimensions provides direct access to physically realistic black hole regimes of mass, charge, and spin, but all nontrivial solutions it generates are constrained by the uniqueness theorem, making most solutions singular or physically pathological.

Several efforts have been made to extend the sigma-model approach to theories with more than four dimensions and broader field content. The general strategy involves assuming $D-3$ Killing vectors in a $D$-dimensional theory, reducing the Einstein action to a sigma model. The viability of this approach depends on the symmetry structure of the resulting sigma model \cite{Cremmer:1999du}.  In some perfect cases, such as vacuum gravity in $D$ dimensions, the symmetry group is SL($D-2,\mathbb{R}$), allowing straightforward extensions of four-dimensional solution-generating techniques, including an Ernst formulation. In contrast, theories like five-dimensional Einstein-Maxwell gravity lack nontrivial hidden symmetries, rendering the sigma-model approach ineffective.  Intermediate scenarios exist where other features, such as the Chern-Simons term in five-dimensional Einstein-Maxwell theory,  yielding minimal supergravity,  restore hidden symmetries, leading to a symmetry group that permits solution generation \cite{Bouchareb:2007ax,Clement:2007qy,Gunaydin:2007qq}. However, even when symmetries exist, generalizing solution-generating techniques is not always straightforward. For instance, while five-dimensional supergravity admits generalizations of Ehlers-Harrison transformations \cite{Bouchareb:2007ax,Clement:2007qy,Gunaydin:2007qq}, an Ernst formulation of the model remains elusive.

Despite these challenges, sigma-model techniques have been successfully employed to construct black geometries in higher-dimensional theories, including supergravity theories, where solving Einstein's equations directly would be prohibitively difficult. These techniques have led to non-extremal charged black holes, rings, strings, and branes \cite{Cvetic:1995kv,Chong:2004na,Giusto:2007fx,Bouchareb:2007ax,Tomizawa:2008qr,Galtsov:2008bmt,Galtsov:2008jjb,Virmani:2012kw,Chow:2014cca}. However, surprisingly few attempts have been made to use this approach to generate non-supersymmetric, smooth and horizonless geometries,  or gravitational solitons,  with the same mass, charge, and spin as non-extremal black holes that could correspond to coherent black hole microstates (see \cite{Giusto:2007tt,Katsimpouri:2014ara,Banerjee:2014hza} for some limited attempts that led to over-rotating solutions). 
\vspace{0.3cm}

In this paper, we propose a novel roadmap for bridging the gap between the supersymmetric and sigma-model approaches by employing sigma-model techniques to generate smooth horizonless solutions in supergravity that share key properties with supersymmetric microstate geometries. The supersymmetric approach has already identified the essential ingredients for such solutions: the only viable mechanism for large families of smooth, horizonless geometries as compact as black holes involves topological deformations along compact extra dimensions, supported by electromagnetic fluxes \cite{Gibbons:2013tqa}. Our goal is to revisit the sigma-model approach in supergravity to incorporate these crucial ingredients while retaining the ability to generate non-supersymmetric solutions.

This approach was initiated by one of the authors in restrictive cases where, in addition to assuming $D-3$ Killing vectors, the solutions were static and axially-symmetric, and many electromagnetic degrees of freedom were truncated. Under these assumptions, the Einstein equations in supergravity decompose into multiple decoupled sectors of Ernst equations \cite{Heidmann:2021cms}, enabling the direct extension of four-dimensional solution-generating techniques to many supergravity theories. This has led to the construction of numerous smooth horizonless geometries with the same mass and charge as non-extremal black holes \cite{Bah:2020ogh,Bah:2021owp,Bah:2021rki,Bah:2022pdn,Heidmann:2022zyd}, including solutions comparable to the Schwarzschild black hole \cite{Bah:2022yji,Bah:2023ows,Dulac:2024cso}. Furthermore, this approach has produced neutral bound states of branes and antibranes in string theory with half the entropy of the Schwarzschild black hole, providing its first microscopic entropy counting \cite{Heidmann:2023kry}. However, the restriction to static spacetimes prohibits access to the Kerr regime, and truncating certain electromagnetic fields limits connections to known supersymmetric smooth geometries.

In this paper, we relax these restrictive assumptions by allowing for rotation and including all possible electromagnetic fluxes. This requires a complete reexamination of the integrable structure and sigma models, a review of existing solution-generating techniques, and the construction of new methods to systematically build smooth horizonless solutions.

To achieve this, we consider $\cN=2$ five-dimensional supergravity coupled to two vector multiplets, also known as the STU model \cite{Cremmer:1984hj}. This theory arises from the dimensional reduction of M-theory on T$^6$ and can be dualized into other supergravity frames. We review the sigma model emerging from this theory, where the SO(4,4) group acts as the target space isometry \cite{Chong:2004na,Galtsov:2008bmt,Galtsov:2008jjb,Chow:2014cca}. We provide the first Ernst formulation of this model, thereby establishing its crucial Ernst structure. This consists of generalizing the Ernst formalism in four dimensions into a \emph{matrix Ernst formalism}, previously analyzed in other contexts \cite{Alekseev:2004zz,Galtsov:1995tnt,Herrera-Aguilar:1997fia,Herrera-Aguilar:1997csq}. We then revisit known solution-generating techniques associated with this model, including SO(4,4) transformations that generate novel solutions by acting on a seed solution with coset model symmetries and the generalized inverse scattering method \cite{Figueras:2009mc,Katsimpouri:2013wka}. Additionally, we introduce new techniques based on the Ernst formalism and compare them with existing methods.

Finally, we demonstrate the relevance and promise of this approach in constructing microstates of non-extremal black holes by explicitly applying some of these techniques. We construct novel static, eight-charge smooth horizonless geometries that possess the same mass and charges as non-extremal black holes of the STU model \cite{Chow:2014cca}, but where the horizon is replaced by a smooth bolt. In M-theory, these eight charges correspond to three M2-brane charges, three M5-brane charges along the T$^6$, as well as Kaluza-Klein monopole (KKm) and momentum (P) charges along the S$^1$. These solutions can be viewed as a generalization of the topological star derived in \cite{Bah:2020ogh,Bah:2020pdz}. However, unlike the topological star, their supersymmetric limit is well-behaved and corresponds to the smooth horizonless $\frac{1}{2}$-BPS center, also known as a \emph{Gibbons-Hawking center}, which serves as the fundamental building block of multicenter bubbling geometries,  a class of microstates of BPS black holes in supergravity \cite{Bena:2007kg,Heidmann:2017cxt,Bena:2017fvm,Warner:2019jll}.  Thus, our new solutions can be viewed as non-BPS extensions of these BPS constituents that are fundamental for constructing coherent black hole microstates in supergravity.

\newpage

The structure of this paper is as follows. In Section \ref{sec:SigmaMod}, we establish our conventions, review the SO(4,4) sigma model of five-dimensional supergravity, and derive the novel Ernst formalism associated with this model. In Section \ref{sec:SolGenTech}, we outline the solution-generating techniques that can be employed to construct exact non-supersymmetric solutions in five-dimensional supergravity. Finally, in Section \ref{sec:NewSol}, we apply some of these techniques to construct novel static, eight-charge microstates of non-extremal black holes and analyze their properties. Readers primarily interested in these new solitons can directly refer to this self-contained section.

\section{Sigma model from five-dimensional Supergravity} 
\label{sec:SigmaMod}

In this section, we derive the three-dimensional sigma model emerging from a timelike toroidal compactification of $\cN=2$ five-dimensional supergravity coupled to two vector multiplets, also known as the STU model or five-dimensional U(1)$^3$ supergravity \cite{Cremmer:1984hj}.  While most of the results are reviewed from previous work \cite{Chong:2004na,Galtsov:2008bmt,Galtsov:2008jjb}, we also provide new contents including a formulation of the equations of motion in terms of \emph{a generalized matrix Ernst formalism}. 

Pure Einstein-Maxwell theory in five dimensions fails to produce three-dimensional sigma-model on a symmetric target space which drastically undermines the application of solution-generating techniques.  Remarkably, adding the Chern-Simons term, as prescribed by five-dimensional supergravity,  which may seem as a complexification, leads to a more symmetric three-dimensional sigma-model with a SO(4,4) group acting as the target space isometry \cite{Chong:2004na,Galtsov:2008bmt,Galtsov:2008jjb,Chow:2014cca}.  This type of integrable structure has been extremely useful, for generating black geometries in supergravity where directly solving Einstein's equations would have been a dramatically difficult algebraic task  \cite{Cvetic:1995kv,Chong:2004na,Giusto:2007fx,Bouchareb:2007ax,Tomizawa:2008qr,Galtsov:2008bmt,Galtsov:2008jjb,Virmani:2012kw,Chow:2014cca}.  

The action of STU supergravity in five dimensions is given by
 \begin{align}\label{eq:L5}
\cS_5 \= & \frac{1}{16 \pi G_5} \int \left( R_5 \star_5 1 \!-\!
\frac12 G_{IJ} dX^I \!\wedge\! \star_5 dX^J \!-\! \frac12G_{IJ} F^I
\!\wedge\! \star_5 F^J \!-\!
\frac{|\epsilon_{IJK}|}{6} F^I \!\wedge\! F^J \!\wedge\! A^K \right)\!\!, \\
G_{IJ}\= &{\rm diag}\left((X^1)^{-2},\ (X^2)^{-2},\
(X^3)^{-2}\right),\qquad F^{I}=dA^{I},\qquad I,J,K=1,2,3.\nn
 \end{align}
where $A^I$ define the three U(1) vector fields,  $X^I$ are the three scalar moduli satisfying the constraint $X^1 X^2 X^3 =1$ and $\epsilon_{IJK}$ is the three-dimensional Levi-Civita tensor.

Note that STU supergravity can be truncated to $\cN=2$ minimal supergravity — equivalent to Einstein-Maxwell theory with a Chern-Simons term — by setting the scalars to trivial values and taking all gauge fields to be equal:
\begin{equation}
\cS_5^\text{min} \=  \frac{1}{16 \pi G_5} \int \left( R_5 \star_5 1 
\!-\! \frac32 \,F \!\wedge\! \star_5 F \!-\! F \!\wedge\! F \!\wedge\! A \right).
\label{eq:L5min}
\end{equation}

\subsection{Three-dimensional reduction}

Following  \cite{Cvetic:1995kv,Chong:2004na,Giusto:2007fx,Bouchareb:2007ax,Tomizawa:2008qr,Galtsov:2008bmt,Galtsov:2008jjb,Virmani:2012kw,Chow:2014cca}, we consider solutions that admit two commuting Killing symmetries,  one timelike and one spacelike, that we parametrize by two coordinates $t$ and $\psi$ respectively.  We decompose the five-dimensional fields such that
\begin{align}
ds_5^2 &\= -\frac{1}{(Z_1 Z_2 Z_3)^\frac{2}{3}} \left(dt+\mu (d\psi + \omega_\psi) + \omega_t \right)^2 +(Z_1 Z_2 Z_3)^\frac{1}{3}\left[ \frac{1}{Z_0} \,(d\psi + \omega_\psi)^2 +Z_0 \,ds_3^2\right]\,, \nn \\
A^I &\= A_t^I  \left(dt+\mu (d\psi + \omega_\psi) + \omega_t\right)+A_\psi^I (d\psi+ \omega_\psi )  + a^I\,, \qquad X^I \= \frac{(Z_1 Z_2 Z_3)^\frac{1}{3}}{Z_I}, \label{eq:Metric&FieldsAnsatz}
\end{align}
where we have defined eleven scalars $(Z_0,Z_I,\mu,A_t^I,A_\psi^I)$ and five one-forms $(\omega_t,\omega_\psi,a^I)$ on the three-dimensional base.  Moreover,  we pick a convention for the Hodge star operator such that $\star_5 1 = Z_0 (Z_1 Z_2 Z_3)^\frac{1}{3}\,\,d\psi\wedge dt \wedge \text{vol}_3$,  where $\text{vol}_3$ is the volume form of the three-dimensional base.

To obtain a purely scalar three-dimensional reduction,  we follow \cite{Bouchareb:2007ax,Tomizawa:2008qr,Galtsov:2008bmt,Galtsov:2008jjb}, to perform dualisation of the one-forms.   For that purpose,  we introduce five Lagrange multipliers ensuring the Bianchi identities of the field strengths: $B_{I}$ are the scalar duals of $a^I$, and $(\Omega_t,\Omega_\psi)$ are the scalar duals of $(\omega_t,\omega_\psi)$.  

For convenience,  we introduce the two-dimensional metric, $h$,  along the $(t,\psi)$ space,  the absolute value of its determinant, $\tau$,  some vectors of $(t,\psi)$ quantities and the two dimentional Levi-Civita symbol $\varepsilon$:
\begin{equation}
\begin{split}
h &\= (Z_1 Z_2 Z_3)^{-\frac{2}{3}} \begin{pmatrix}
-1 & -\mu  \\
-\mu &  \frac{Z_1 Z_2 Z_3}{Z_0} - \mu^2 
\end{pmatrix}\,,\qquad \tau \= - \det h \= \frac{1}{Z_0(Z_1 Z_2 Z_3)^{\frac{1}{3}}}\,,\\
 \omega &\= \begin{pmatrix}
\omega_t \\
\omega_\psi
\end{pmatrix}\,,\qquad \Omega \= \begin{pmatrix}
\Omega_t \\
\Omega_\psi
\end{pmatrix}\,,\qquad \cA^I\= \begin{pmatrix}
A_t^I \\
A_\psi^I + \mu A_t^I
\end{pmatrix}\,,\qquad  \varepsilon\= \begin{pmatrix}
0 && -1 \\
1 && 0
\end{pmatrix}\,.
\end{split}
\end{equation}
The dualization of the one-forms are given by\footnote{They are derived from the Maxwell equations $d\left(G_{IJ} \star_5 F^{J} - \frac{|\epsilon_{IJK}|}{2} A^J \wedge F^K \right)=0$.}
\begin{equation}
\begin{split}
da^I + {\cA^I}^T \,d\omega &\= \tau^{-1} G^{IJ} \star \cB_J \,,\qquad \cB_I \equi dB_{I} +\frac{|\epsilon_{IJK}|}{2} \,d{\cA^J}^T \varepsilon \cA^K, \\
 \tau h d\omega &\= \star \cV \,,\qquad \cV \= d\Omega- \left(dB_I + \frac{|\epsilon_{IJK}|}{6}  \,d{\cA^J}^T \varepsilon \cA^K  \right) \cA^I \,,
 \end{split}
 \label{eq:ScalarDuals}
\end{equation}
where $\star$ is the Hodge star operator in the three-dimensional space,  and $G^{IJ}= {\rm diag}\left((X^2 X^3)^{-2},\ (X^1 X^3)^{-2},\ (X^1 X^2)^{-2}\right)$ is the inverse of $G_{IJ}$.

Now,  we can recast the three-dimensional reduced action in terms of 16 scalars,  $\varphi^a = (Z_0,Z_I,\mu,A_t^I,A_\psi^I,B_I,\Omega_t,\Omega_\psi)$:
 \begin{align}\label{eq:L3}
\cS_3 \=  \frac{1}{16 \pi G_3} \int \left( R_3 \star 1 \-
 \cG_{ab} \partial_\mu \varphi^a \partial^\mu \varphi^b \,\star 1 \right)\,.
 \end{align}
The target space of the coset model is given by the line element $dl^2=\cG_{ab}d\varphi^a d\varphi^b$:
 \begin{align}
 dl^2 =&\frac14 \mathrm{Tr} \left(
h^{-1} dh h^{-1} dh \right)+ \frac14\tau^{-2} d\tau^2 + \frac12 G_{IJ}(dX^IdX^J+d{{\cA^I}^T}h^{-1}
d\cA^J)-\frac12\tau^{-1}G^{IJ}\cB_I \cB_J\nn\\
  & - \frac12\tau^{-1} \cV^T h^{-1} \cV.\label{eq:TS_metric}
 \end{align}
 As discussed in \cite{Galtsov:2008bmt,Galtsov:2008jjb, Chow:2014cca},  this three-dimensional sigma model is invariant under the action of the 28-parametric SO(4, 4) isometry group.
 
 \subsection{Matrix representation}
 
 In \cite{Galtsov:2008bmt,Galtsov:2008jjb},  a $8\times 8$ matrix representation of the sigma model that makes the SO(4, 4) symmetry manifest has been derived.  In this section,  we use a slightly different form that will make the Ernst formulation explicit.\footnote{The change from their matrix to ours is
 \begin{equation}
\cM^\text{ours} = \cJ \cM^\text{theirs} \cJ,\quad \text{where}\quad  \cJ = \begin{pmatrix}
 J & 0 \\
 0 & \mathbb{I}_4
 \end{pmatrix},\qquad J \equiv \begin{pmatrix}
 0 & 0 &0 & 1 \\
 0 & 0& 1&0 \\
  0 & 1& 0&0 \\
   1 & 0& 0&0 
 \end{pmatrix}\label{eq:ft1}
 \end{equation} }
 The matrix representation of the coset provides a compact formulation of the three-dimensional action \eqref{eq:L3}:
  \begin{equation}
  \label{eq:L3Matrix}
\cS_3 \=  \frac{1}{16 \pi G_3} \int \left( R_3 \star 1+ \frac{1}{8} \text{Tr} \left[d\cM^{-1} \wedge \star d\cM \right] \right)\,.
 \end{equation}
The matrix representation of the coset ${\cal M}$ decomposes into
$4\times 4$ block matrices $P=P^T$ and $Q^T =-Q$ as follows
\begin{equation}
\cM \equiv \begin{pmatrix} 
P && P Q \\
-Q P && P^{-1} - Q P Q
\end{pmatrix}\,,\qquad \cM^{-1} \= \begin{pmatrix} 
P^{-1} - Q P Q && -Q P \\
P Q && P
\end{pmatrix},
\label{eq:MatrixMDef}
\end{equation}
where 
\begin{align}
Q & \equi \begin{pmatrix} 
0 &&  \widetilde{A}_\psi^2 && A_t^2&&- B_1 + \frac{ A_t^3  \widetilde{A}_\psi^2 -  A_t^2  \widetilde{A}_\psi^3}{2} \\
* && 0&& B_3 + \frac{ A_t^2  \widetilde{A}_\psi^1 -  A_t^1  \widetilde{A}_\psi^2}{2} && - \Omega_\psi +  \widetilde{A}_\psi^2 B_2 +\frac{ A_t^3  \widetilde{A}_\psi^1  \widetilde{A}_\psi^2-2A_t^2  \widetilde{A}_\psi^1  \widetilde{A}_\psi^3+A_t^1  \widetilde{A}_\psi^2  \widetilde{A}_\psi^3}{6} \\
* && * && 0 && - \Omega_t + A_t^2 B_2 -\frac{ A_t^2 A_t^3  \widetilde{A}_\psi^1-2A_t^1 A_t^3  \widetilde{A}_\psi^2+A_t^1 A_t^2   \widetilde{A}_\psi^3}{6} \\
* && * && * && 0
\end{pmatrix} \\
 P& \equi \begin{pmatrix} 
\frac{Z_2}{Z_1}-\Phi^T \Lambda \Phi  && \Phi^T \Lambda \Psi  \\
\Psi^T \Lambda \Phi && -\Psi^T \Lambda \Psi 
\end{pmatrix},
\end{align}
where we have defined $\widetilde{A}^I_\psi \equiv A_\psi^I +\mu A_t^I$ and
\begin{equation}
\Psi \= \begin{pmatrix} 
-1&&0&&0\\
0 && 1&& 0 \\
A_t^3 && - \widetilde{A}_\psi^3&& -1
\end{pmatrix}, \qquad \Phi \=  \begin{pmatrix} 
 -\widetilde{A}^1_\psi \\
A_t^1 \\
B_2 +\frac{A_t^3 \widetilde{A}_\psi^1-A_t^1 \widetilde{A}_\psi^3}{2}
\end{pmatrix}\,,\quad \Lambda \= \begin{pmatrix} 
\frac{Z_0(Z_1 Z_2)^\frac{2}{3}}{Z_3^\frac{1}{3}}\,h&&0\\
0 && Z_0Z_3
\end{pmatrix}.
\end{equation}

One can check that the coset matrix $\cM$ indeed belongs to SO(4,4) as
\begin{equation}
\cM^{-1}   \= \begin{pmatrix}
0 && \mathbb{I}_4 \\
\mathbb{I}_4 && 0
\end{pmatrix} \,  \cM^T \, \begin{pmatrix}
0 && \mathbb{I}_4 \\
\mathbb{I}_4 && 0
\end{pmatrix} \,,\qquad \det \cM = 1\,,
\label{eq:SO(4,4)M}
\end{equation}
with the additional constraint that $\cM$ is symmetric,  $\cM^T =\cM$.

The equations of five-dimensional supergravity admit a compact form in terms of $\cM$:
\begin{equation}
d\left(\cM^{-1} \star d\cM \right) \= 0\,,\qquad (R_3)_{\mu \nu} \= - \frac{1}{8} \text{Tr}\left[ \partial_\mu \cM^{-1} \partial_\nu \cM \right].
\label{eq:EOMMatrixM}
\end{equation}
From the first equation,  we can introduce the dual matrix-valued one-form that will contain information about the one-forms of the five-dimensional solution,  $(\omega_t,\omega_\psi,a^I)$,
\begin{equation}
d\cN \equi \cM^{-1} \star d\cM\,.
\label{eq:NMatrixDef}
\end{equation}

Finally,  the action \eqref{eq:L3Matrix} is manifestly invariant under SO(4,4) transformation:\footnote{Only left and right transformations are considered to preserve $\cM^T = \cM$.}
\begin{equation}\label{eq:SOTransformation}
\cM \,\to\,  g^{T} \cM g\,,\qquad \cN \,\to\,  g^{-1} \cN g \,,\qquad g\in \text{SO(4,4)}.
\end{equation}

\subsection{Ernst formulation}
\label{sec:ErnstFormalism}

The Ernst formalism was originally developed in four-dimensional Einstein-Maxwell theory \cite{Ernst:1967wx}, sparking a wide array of solution-generating techniques in classical gravity, such as inverse scattering methods \cite{Belinsky:1971nt,Belinsky:1979mh}, Bäcklund-Ehlers-Harrison transformations \cite{harrison_new_1968,Geroch:1970nt,PhysRevLett.41.1197,Harrison:1980fr,Alekseev:2020bqu}, and the Sibgatullin method \cite{Manko_1993}.

In four dimensions, the Ernst formalism introduces two complex potentials: one governing the electromagnetic fields,  $\Phi$,  and the other encapsulating the gravitational field and spin $\cE$. Extending this framework to higher dimensions is nontrivial, and only a limited number of higher-dimensional theories admit an Ernst formulation. These include specific compactifications of heterotic string theory \cite{Herrera-Aguilar:1997csq} and Einstein-Maxwell-dilaton-axion theories \cite{Galtsov:1994pd,Galtsov:1997jrl}. Such generalizations typically require transitioning from scalar Ernst potentials to matrix Ernst potentials, leading to what is termed the \emph{generalized matrix Ernst formalism} \cite{Alekseev:2004zz}.

From the previous section,  rewriting the sigma model in terms of the matrix $\cM$ as in \eqref{eq:MatrixMDef} enables a direct reformulation using Ernst equations, marking the first derivation of the matrix Ernst structure in five-dimensional STU supergravity. We introduce the $4 \times 4$ real-valued matrix Ernst potential $\cE$, defined in terms of the matrices $P$ and $Q$ as follows:
\begin{equation}
\cE \equi P^{-1} - Q\,.
\label{eq:ErnstPotDef}
\end{equation}
In contrast to the four-dimensional Ernst formalism, where the Ernst potential is complex-valued, the symmetric part ($P^{-1}$) and skew-symmetric part ($Q$) of $\cE$ play analogous roles to the real and imaginary components of the Ernst potential in four dimensions.

Notably, unlike the four-dimensional case, there is only one Ernst potential, and the potential $\Phi$ associated to the electromagnetic fields is absent. Instead, the U(1) gauge fields of STU supergravity are incorporated directly into the matrix potential $\cE$.  This could have been expected from the fact that $\cE$ has already 16 scalar degrees of freedom matching the 16 scalars of STU supergravity.

Consequently, the equations of motion \eqref{eq:EOMMatrixM} reduce to the vacuum matrix Ernst equations:
\begin{equation}
d \star d\cE \= 2 d\cE \wedge \left(\left(\cE+\cE^T \right)^{-1} \star d \cE\right)\,,\qquad (R_3)_{\mu \nu} \= \text{Tr}\left[ (\partial_\mu \cE) \left(\cE+\cE^T \right)^{-1} ( \partial_\nu \cE^T) \left(\cE+\cE^T \right)^{-1}  \right].
\label{eq:EOMErnst}
\end{equation}

As for the coset matrix $\cM$,  it is relevant to introduce one-form potentials
\begin{equation}\label{eq:NblockMatrixDef}
d\cN_1 \= P \star d Q P \,,\qquad d\cN_2 \= -\cE d\cN_1- 2\star d\cE (\cE+\cE^T)^{-1}\,,\qquad d\cN_3 \= \cE d\cN_1 \cE^T +d\cN_2 \cE^T -\cE d\cN_2^T,
\end{equation}
and the one-form matrix,  $\cN$, \eqref{eq:NMatrixDef} is given by
\begin{equation}
\label{eq:NMatrixDecomp}
\cN \= \begin{pmatrix}
\cN_2 && \cN_3 \\
-\cN_1 &&  -\cN_2^T
\end{pmatrix}.
\end{equation}

This decomposition facilitates the extension of solution-generating techniques, which will be explored in a future section.

\subsection{Cohomogeneity-two solutions}
\label{sec:Coho2}

We can further assume that the solutions admit an extra spacelike isometry,  so that the solutions are cohomogeneity-two solutions,  meaning it depends on two spacetime coordinates. This allows to write the three-dimensional base as
\begin{equation}
ds_3^2 \= e^{2\nu}\left(d\rho^2+dz^2 \right)+\rho^2 d\phi^2,
\label{eq:3DBase}
\end{equation}
where $(\rho,z,\phi)$ are commonly called the Weyl-Papapetrou coordinates,  $e^{2\nu}$ is the conformal factor,  and all fields depend on $(\rho,z)$ only.

For most of the solutions of interest,  the one-form gauge fields $(a^I,\omega_t,\omega_\psi)$ have only a component along $d\phi$: 
\begin{equation}
\omega_t \to \omega_{t\phi} \,d\phi \,,\qquad\omega_{\psi} \to \omega_{\psi\phi}  \,d\phi \,,\qquad a^I \to A_\phi^I \,d\phi,
\end{equation}
where $\omega_{t\phi}$,  $\omega_{\psi\phi}$ and $A_\phi^I$ are now gauge potentials.  The three-dimensional Hodge operator used in the previous sections simplifies for one-forms $X=X_\phi d\phi$ and scalars $Y$ such as
\begin{equation}
\star dX \to \rho^{-1} \star_2 dX_\phi\,,\qquad  \star dY  \to \rho \star_2 dY\wedge d\phi\,,
\end{equation}
where $\star_2$ is the Hodge operator in the flat $(\rho,z)$ space.  This allows to rewrite the equations of motion \eqref{eq:EOMMatrixM} and \eqref{eq:EOMErnst} in terms of the Laplacian and gradient in flat space in cylindrical coordinates for $\phi$-independent functions,  $\Delta=\frac{1}{\rho} \partial_\rho \left(\rho \,\partial_\rho \right) +\partial_z^2$ and $\nabla=(\partial_\rho,\partial_z)$,
\begin{align}
\nabla \left(\rho \cM^{-1}\nabla \cM \right) \= 0,\qquad \Delta \cE \= 2 \nabla \cE. \left(\cE+\cE^T \right)^{-1} .\nabla \cE\,,
\label{eq:2dEquations}
\end{align}
and the equation governing the conformal factor $e^{2\nu}$ is obtained from \eqref{eq:EOMMatrixM} and \eqref{eq:EOMErnst} using $\partial_z \nu \= \rho (R_3)_{\rho z}$ and $2\partial_\rho \nu \= \rho \left((R_3)_{\rho \rho}-(R_3)_{z z} \right)$.

\subsection{Extracting the supergravity fields}

In the previous sections,  we have shown that the STU sigma model is captured either by the coset matrix $\cM$ or the matrix Ernst potential $\cE$.  In this section, we show how one can retrieve the fields of the five-dimensional solutions from these quantities.

\subsubsection{From the coset matrix}

The three-dimensional scalars can be extracted from the coset matrix $\cM$ \eqref{eq:MatrixMDef} such as
\begin{align}
&Z_0 \=\sqrt{\cM_{42}^2 - \cM_{44} \cM_{22}}\,,\qquad Z_1 Z_2 \= -\cM_{33} - Z_0^{-2} \left(\cM_{22} \cM_{43}^2 + \cM_{44} \cM_{32}^2 - 2 \cM_{32} \cM_{42} \cM_{43}\right)
 ,\nn\\ 
&Z_3  \= -\frac{\cM_{44}}{\sqrt{\cM_{42}^2 - \cM_{44} \cM_{22}}},\qquad \mu \= \frac{\cM_{32} \cM_{44} - \cM_{42} \cM_{43}}{\cM_{42}^2 - \cM_{44} \cM_{22}},\qquad  A_t^3 \= -\frac{\cM_{42}}{\cM_{44}}\,,\nn \\
&A_t^1 \=  \frac{ \cM_{41} \cM_{43} -\cM_{31} \cM_{44} + Z_0^{-2} \left(\cM_{44} \cM_{21} - \cM_{42} \cM_{41}\right) \left(\cM_{42} \cM_{43} - \cM_{32} \cM_{44}\right)}
{ \cM_{33}\cM_{44} + Z_0^{-2} \cM_{44}\left(\cM_{22} \cM_{43}^2 + \cM_{44} \cM_{32}^2 - 2 \cM_{32} \cM_{42} \cM_{43}\right)} \,,\nn \\
&A_t^2 \=  \frac{ \cM_{54} \cM_{43} -\cM_{53} \cM_{44} + Z_0^{-2} \left(\cM_{44} \cM_{52} - \cM_{42} \cM_{54}\right) \left(\cM_{42} \cM_{43} - \cM_{32} \cM_{44}\right)}
{ \cM_{33}\cM_{44} + Z_0^{-2} \cM_{44}\left(\cM_{22} \cM_{43}^2 + \cM_{44} \cM_{32}^2 - 2 \cM_{32} \cM_{42} \cM_{43}\right)}  ,  \label{eq:ExtractingFields}\\
& \frac{Z_2}{Z_1} - Z_1 Z_2(A_t^1)^2 \= \cM_{11} +Z_0^{-2} \left( \cM_{22} \cM_{41}^2 + \cM_{44} \cM_{21}^2 - 2 \cM_{21} \cM_{42} \cM_{41} \right)\,,\nn \\
& A_\psi^1 \=\frac{\cM_{44} \cM_{21} - \cM_{41} \cM_{42}}{\cM_{42}^2 - \cM_{44} \cM_{22}},\quad A_\psi^2 \= \frac{\cM_{44} \cM_{52} - \cM_{42} \cM_{54}}{\cM_{42}^2 - \cM_{44} \cM_{22}},\quad A_\psi^3 \= \frac{\cM_{32} \cM_{42} - \cM_{22} \cM_{43}}{\cM_{42}^2 - \cM_{44} \cM_{22}}. \nn
\end{align}
It will be also useful to express the metric quantity, $\cI_4$, that intervenes in the four-dimensional reduction of the solution along the $\psi$ direction:
\begin{equation}
\cI_4 \equi Z_0 Z_1 Z_2 Z_3  - \mu^2 Z_0^2 \=\cM_{44}\cM_{33}-\cM_{43}^2.
\label{eq:QuarticDef}
\end{equation}
Finally,  the five one-forms are derived from the dual $\cN$ matrix \eqref{eq:NMatrixDef}.  We find:
\begin{equation}
\omega_t \= -\cN_{74}\,,\qquad \omega_\psi \= -\cN_{64}\,,\qquad a^1 \= -\cN_{54}\,,\qquad a^2 \= -\cN_{14}\,,\qquad a^3 \=  \cN_{63}\,.
\end{equation}

\subsubsection{From the Ernst potential}

One can similarly extract the three-dimensional scalars from the matrix Ernst potential \eqref{eq:ErnstPotDef}.  We have identical expressions as in \eqref{eq:ExtractingFields} by simply replacing
\begin{equation}
\cM_{ij} \= P_{ij} \= 2\left((\cE+\cE^T)^{-1}\right)_{ij},\qquad i,j=1,2,3,4,
\end{equation}
for all scalars except $A_t^2$ and $A_\psi^2$ that involve $\cM$ components beyond the first $4\times 4$ block.  For those two, we have:
\begin{equation}
A_t^2 \= Q_{13} \= \frac{1}{2}(\cE^T - \cE)_{13}\,,\qquad A_\psi^2 + \mu A_t^2 \= Q_{12} \= \frac{1}{2}(\cE^T - \cE)_{12}.
\end{equation}
Moreover, working with the Ernst potential enables us to extract the fields using $P$ and its inverse $\widetilde{P} \equiv P^{-1} = \frac{1}{2}(\cE + \cE^T)$, which, for certain fields, results in simpler expressions. For example:
\begin{equation}
Z_2 \= \frac{1}{\sqrt{\widetilde{P}_{31}^2-\widetilde{P}_{33}\widetilde{P}_{11}}},\qquad Z_1 \= \widetilde{P}_{11}\,Z_2\,,\qquad A_t^1 \= \frac{\widetilde{P}_{31}}{\widetilde{P}_{11}}\,.
\end{equation}

Finally,  the five one-forms are derived from the dual $\cN_I$ matrices \eqref{eq:NblockMatrixDef}.  We find:
\begin{equation}
\omega_t \= (\cN_1)_{34}\,,\qquad \omega_\psi \= (\cN_1)_{24}\,,\qquad a^1 \= (\cN_1)_{14}\,,\qquad a^2 \= -(\cN_2)_{14}\,,\qquad a^3 \=  (\cN_1)_{23}\,.
\end{equation}

\subsection{Dual supergravity frames}

While we have presented the sigma model arising from five-dimensional supergravity, the same sigma model could emerge from various other dual supergravity frames. In this section, we review some of the relevant ones.

\subsubsection{Eleven-dimensional supergravity}

Five-dimensional $\mathcal{N}=2$ supergravity coupled to two vector multiplets can be viewed as the compactification of eleven-dimensional supergravity on a rigid T$^6 =$T$^2 \times$T$^2 \times$T$^2$ (see, for instance, \cite{Emparan:2006mm,DallAgata:2010srl}):
\begin{align}
ds_{11}^2 \= ds_5^2 \+ \sum_{I=1}^3 X^I \left(dx_{2I-1}^2+ dx_{2I}^2 \right) \,,\qquad F_4 \= dA_3= \sum_{I=1}^3 F^I \wedge dx_{2I-1}\wedge dx_{2I}\,,
\label{eq:MTheoryEmbed}
\end{align}
where $F_4$ is the four-form field strength,  $x_i$ parametrize the T$^6$ coordinates and $(ds_5^2,X^I,F^I=dA^I)$ are the five-dimensional fields given in \eqref{eq:Metric&FieldsAnsatz}.  

One advantage of the eleven-dimensional embedding is that the action and Einstein equations take a more compact form than their five-dimensional supergravity counterparts. The action is given by
 \begin{align}\label{eq:L11}
\cS_{11} \= & \frac{1}{16 \pi G_{11}} \int \left( R\, \star_{11} 1 \!-\!
\frac12 F_4 \!\wedge\! \star_{11} F_4 \!-\!
\frac{1}{6} F_4 \!\wedge\! F_4 \!\wedge\! A_3 \right)\!\!, \qquad G_{11} \= 16 \pi^7 l_p^9\,,
 \end{align}
and the Einstein-Maxwell equations are
 \begin{equation}
 d \star_{11} F_4 \- \frac{1}{2} F_4 \wedge F_4 \= 0\,,\quad R_{\mu \nu} \= \frac{1}{12} \left(F_{4\mu\alpha\beta\gamma}F_{4\nu}^{\,\,\, \,\alpha\beta\gamma}-\frac{1}{12} \,g_{\mu\nu}\, F_{4\alpha\beta\gamma\delta}F_{4}^{\,\,\alpha\beta\gamma\delta}\right).
 \end{equation}

\subsubsection{Type IIB supergravity}
\label{sec:typeIIB}

Another interesting supergravity framework is the D1-D5-P frame in type IIB supergravity on T$^4 \times$ S$^1$. This frame is obtained through a dimensional reduction from eleven dimensions followed by a series of T-dualities (see, for instance, \cite{Bena:2008dw}).
The type IIB metric is 
\begin{align}
ds_\text{IIB}^2 \= \sqrt{X^3}\,ds_5^2 \+(X^3)^{-\frac{3}{2}} (dy+A^3)^2+X^2 \sqrt{X^3} (dx_1^2+dx_2^2+dx_3^2+dx_4^2) \,,
\label{eq:TypeIIBsol}
\end{align}
where $x_i$ and $y$ parametrize T$^4$ and S$^1$, respectively, and $(ds_5^2, X^I, A^I)$ are the five-dimensional fields given in \eqref{eq:Metric&FieldsAnsatz}.

The other non-vanishing fields include the dilaton and the RR three-form field strength:
\begin{equation}
e^{2\Phi} \= (X^2)^2 X^3 \= \frac{Z_1}{Z_2}\,,\qquad F^{(3)} \= - (X^2)^{-2}\,\star_5 F^2 - F^1 \wedge (dy-A^3)\,,
\end{equation}
where $\star_5$ is the Hodge star operator in the five-dimensional spacetime.

\subsubsection{Four-dimensional $\cN=2$ supergravity}

When the $\psi$ direction defines a compact circle, the dimensional reduction of $\mathcal{N}=2$ five-dimensional supergravity coupled to two vector multiplets leads to four-dimensional STU supergravity, which consists of three complex scalars and four U(1) gauge fields (see for instance \cite{DallAgata:2010srl}).

To perform the dimensional reduction of the five-dimensional ansatz \eqref{eq:Metric&FieldsAnsatz}, it is first rewritten as a fibration along $\psi$:
\begin{align}
ds_5^2 &\= \frac{\cI_4}{Z_0^2(Z_1 Z_2 Z_3)^\frac{2}{3}} \left(d\psi-\frac{\mu Z_0^2}{\cI_4} (dt + \omega_t) + \omega_\psi \right)^2 + \frac{Z_0(Z_1 Z_2 Z_3)^\frac{1}{3}}{\sqrt{\cI_4}} \left[- \frac{1}{\sqrt{\cI_4}} \left( dt+\omega_t\right)^2 +\sqrt{\cI_4} \,ds_3^2\right], \nn \\
A^I &\= (A_\psi^I+\mu A_t^I)  \left(d\psi-\frac{\mu Z_0^2}{\cI_4} (dt + \omega_t) + \omega_\psi \right)+\frac{1}{\cI_4}\left(Z_0 Z_1 Z_2 Z_3 A^I_t+\mu Z_0^2 A_\psi^I \right)(dt+\omega_t) + a^I, \nn
\end{align}
where $\cI_4$ is defined in \eqref{eq:QuarticDef}.

Thus, the corresponding four-dimensional STU supergravity solution is given by
\begin{equation} 
\begin{split}
ds_4^2 & \= - \frac{1}{\sqrt{\cI_4}} \left( dt+\omega_t\right)^2 +\sqrt{\cI_4} \,ds_3^2\,,\qquad z_I \= A^{I}_\psi + \mu A^{I}_t + i \, \frac{\sqrt{\cI_4}}{Z_0 Z_I}\,, \\
A_{(4d)}^0 & \= \frac{\mu Z_0^2}{\cI_4} \left( dt+\omega_t \right) - \omega_\psi\,,  \qquad A_{(4d)}^I \= \frac{1}{\cI_4}\left(Z_0 Z_1 Z_2 Z_3 A^I_t+\mu Z_0^2 A_\psi^I \right)(dt+\omega_t) + a^I\,,
\end{split}
 \label{eq:4dProfile}
\end{equation}
where $z_I$ are the three scalars and $A^I_{(4d)}$ are the four gauge fields.

\subsection{Conserved charges}

In general, the conserved charges of a solution in $\cN=2$ five-dimensional supergravity depend on its asymptotic structure. Since this paper focuses on constructing solutions asymptotic to $\IR^{1,3} \times \text{S}^1$, where the $\psi$ circle is a compact dimension with periodicity $\psi = \psi + 2\pi R_{\psi}$, we restrict our analysis to such solutions. 

For an asymptotically $\IR^{1,3} \times \text{S}^1$ solution, there are ten independent conserved charges at first order in the asymptotic radial expansion around Minkowski space: the ADM mass $M_\text{ADM}$, angular momentum $J$, three electric charges $Q_I$, and three magnetic charges $P_I$ associated with the gauge fields, along with a Kaluza-Klein monopole charge $P_0$ and a momentum charge $Q_0$ in the Kaluza-Klein vector.

The electromagnetic charges are defined by the following integrals of the five-dimensional fields \eqref{eq:Metric&FieldsAnsatz} at large distances:\footnote{The sign difference between the Kaluza-Klein charges and the gauge field charges arises because the reduction to four dimensions \eqref{eq:4dProfile} introduces an overall minus sign.}
\begin{equation}
\begin{split}
P_0 &\= \frac{-1}{4\pi} \int_{S^2} d\omega_\psi,  \qquad P_I \= \frac{1}{4\pi} \int_{S^2} dA^I,\\
 Q_0 &\= \frac{1}{8\pi R_{\psi}} \int_{S^2\times S^1}\star_5 d\left(\frac{\mu Z_0^2}{\cI_4} \,dt \right),\qquad Q_I \= \frac{1}{8\pi R_{\psi}} \int_{S^2\times S^1} (X^I)^{-2} \star_5 dA^I - \frac{|\epsilon_{IJK}|}{2}  A^J \wedge dA^K\,.
\end{split}
\end{equation}
In M-theory on S$^1\times$T$^2 \times$T$^2 \times$T$^2$ \eqref{eq:MTheoryEmbed},  $P_0$ and $Q_0$ correspond to the KKm and P-momentum charges along the S$^1$ respectively,  $P_I$ to the three M5-brane charges along the three S$^1\times$T$^2 \times$T$^2$,  and $Q_I$ to the three M2-brane charges along the three T$^2$.   

When all fields are well-behaved asymptotically, meaning that the $Z$'s and $X^I$ approach $1$ and the electric gauge fields approach $0$, the charges, mass, and spin can be extracted from the asymptotic expansions of the five-dimensional fields \eqref{eq:Metric&FieldsAnsatz}:
\begin{equation}
\begin{split}
\omega_\psi& \sim P_0 \cos \theta d\phi , \qquad a^I \sim - P_I \cos \theta d\phi , \qquad A^I_t \sim - \frac{Q_I}{r}\,,\qquad \mu \sim - \frac{Q_0}{r}\,,\\
 \cI_4 &\sim 1+ \frac{4G_4M_\text{ADM}}{r},\qquad \omega_t \sim 2J \frac{\sin^2 \theta}{r} d\phi,
\end{split}
\label{eq:conservedchargesGen}
\end{equation}
where $(r, \theta, \phi)$ are the spherical coordinates of the asymptotic flat $\IR^3$ space and $G_4$ is the four-dimensional Newton constant,  $G_4 = G_5/(2\pi R_{\psi})$.

\section{Solution-generating techniques}
\label{sec:SolGenTech}

In the previous section,  we have shown that $\cN=2$ five-dimensional supergravity coupled to two vector multiplets decomposes into a three-dimensional sigmal model with a SO(4,4) group acting as the target space isometry.  The sigma model is captured by the $8\times8$ coset matrix $\cM$ \eqref{eq:MatrixMDef},  and its one-form dual $\cN$ \eqref{eq:NMatrixDef},  and the equations of motion are invariant under SO(4,4) transformation \eqref{eq:SOTransformation}.

These transformations allow to transform a well-known seed solution to a new solution with a different field content.  This solution-generating technique has been extensively used to generate non-supersymmetric black holes,  black strings, and black rings in supergravity from known black seed solutions (see \cite{Cvetic:1995kv,Chong:2004na,Giusto:2007fx,Bouchareb:2007ax,Tomizawa:2008qr,Galtsov:2008bmt,Galtsov:2008jjb,Virmani:2012kw,Chow:2014cca} as a non-exhaustive list).  In order to apply the same solution-generating technique to the construction of non-supersymmetric smooth horizonless geometries by starting with a smooth seed solution,  we review the technique in this section.

Moreover,  we will use the Ernst formulation of the sigma model to build similar methods that could yield interesting solution-generating techniques.  Those are the generalization Ehlers transformations of the Ernst formalism \cite{harrison_new_1968,Geroch:1970nt,PhysRevLett.41.1197,Harrison:1980fr,Alekseev:2020bqu}.  

Finally, for completeness, we will also review the generalization of the Belinsky-Zakharov inverse scattering method \cite{Belinsky:1971nt,Belinsky:1979mh,Belinski:2001ph} to the STU supergravity sigma model, as derived in \cite{Figueras:2009mc,Katsimpouri:2013wka}. While this method will not be applied in this paper, it is a promising solution-generating technique that could enable the construction of non-extremal, smooth horizonless geometries composed of multiple bubbles in spacetime, capable of carrying arbitrary charges and spin.

\subsection{SO(4,4) transformations}

As discussed in the previous section,  the sigma model is invariant under the transformation \eqref{eq:SOTransformation}.  Thus, from a seed solution characterized by the matrices $(\cM_0,\cN_0)$,  the transformed fields 
\begin{equation}\label{eq:TNTransfo}
\cM \,=\,  g^{T} \cM_0\, g\,,\qquad \cN \,=\,  g^{-1} \cN_0\, g \,,\qquad g\in \text{SO(4,4)},\qquad ds_3^2 \= ds_{30}^2\,,
\end{equation}
form a new solution of five-dimensional supergravity where $ds_3^2$ and $ds_{30}^2$ are the three-dimensional base of the solutions that remain invariant under transformation.  

In this section, we first decompose the $\text{SO}(4,4)$ group at the level of its Lie algebra, before focusing on the $\text{SO}(4,4)$ transformations that preserve the asymptotic flatness of the seed solution.

\subsubsection{$\mathfrak{so}(4,4)$ generators}

We construct the generators of the Lie Group SO(4,4) from its Lie algebra,  $\mathfrak{so}(4,4)$, which is a $28$ dimensional vector space generated by the  the $8\times 8$ matrices
\begin{equation}\label{Tbasis}
\mathcal{T}=\left\{\cH_i, \cP_{\pm I}, \cW_{\pm I}, \cZ_{\pm I}, \cO_{\pm p}, \cX_\pm \right\},
\end{equation}
with $i=1,2,3,4,I=1,2,3, p=1,2$.  The elements of SO(4,4) can be obtained by the exponential map:
\begin{equation}
g \= e^\mathfrak{g}\,\in\, \text{SO}(4,4)\,,\qquad\text{where}\quad \mathfrak{g}\,\in\, \mathfrak{so}(4,4)\,.
\end{equation}

The $H$-generators are the $4$ Cartan generators of $\mathfrak{so}(4,4)$, the $\cP_{+ I}, \cW_{+ I}, \cZ_{+ I}, \cO_{+ p}, \cX_+$ are $12$-positive root generators and $\cP_{- I}, \cW_{- I}, \cZ_{- I}, \cO_{- p}, \cX_-$ are $12$-negative root generators (see \cite{Galtsov:2008bmt,Galtsov:2008jjb,Berkooz:2008rj} for more details). The positive and negative root generators are related to each other by matrix transpose,  $\cY_{-a} \= (\cY_{+a})^T$.

First,  we introduce  the $8\times 8$ matrices $E_{ij}, \ i,j=1,2,\cdots, 8$ with 1 in the $(i,j)$ component and zero elsewhere:
\begin{equation}
(E_{ij})_{kl} \= \delta_{ik} \delta_{jl}\,,\qquad k,l \= =1,2,\cdots, 8.
\end{equation}
The $28$ generators of $\mathfrak{so}(4,4)$ can be expressed in the following compact form:\footnote{The generators have been derived from  \cite{Galtsov:2008bmt,Galtsov:2008jjb} by applying the map \eqref{eq:ft1}.}
\begin{itemize}
\item[•]\underline{The $H$-generators:}
\begin{equation}
\cH_1\equi E_{44}-E_{88},\qquad \cH_2\equi E_{33}-E_{77},\qquad \cH_3\equi E_{22}-E_{66},\qquad \cH_4\equi E_{11}-E_{55}.
\end{equation}

\item[•]\underline{The $\cP$-generators:}
\begin{equation}
\begin{split}
\cP_{+1}&\equi E_{45}-E_{18},  \qquad \cP_{+2}\equi E_{41}-E_{58},  \qquad \cP_{+3}\equi E_{27}-E_{36},  \\
\cP_{-1} &\equi E_{54}-E_{81},\qquad \cP_{-2}\equi E_{14}-E_{85},\qquad  \cP_{-3}\equi E_{72}-E_{63}.
\end{split}
\end{equation}

\item[•]\underline{The $\cW$-generators:}
\begin{equation}
\begin{split}
\cW_{+1} &\equi E_{56}-E_{21},   \qquad \cW_{+2}\equi E_{16}-E_{25},   \qquad \cW_{+3}\equi E_{43}-E_{78},  \\
\cW_{-1} &\equi E_{65}-E_{12},  \qquad \cW_{-2}\equi E_{61}-E_{52}, \qquad \cW_{-3}\equi E_{34}-E_{87}.
\end{split}
\end{equation}

\item[•]\underline{The $\cZ$-generators:}
\begin{equation}
\begin{split}
\cZ_{+1}\equi  E_{57}-E_{31},   \qquad \cZ_{+2}\equi E_{17}-E_{35},  \qquad \cZ_{+3}\equi E_{68}-E_{42},  \\
\cZ_{-1}\equi  E_{75}-E_{13}, \qquad  \cZ_{-2}\equi E_{71}-E_{53},  \qquad  \cZ_{-3}\equi E_{86}-E_{24}.
\end{split}
\end{equation}

\item[•]\underline{The $\cO$-generators:}
\begin{equation}
\cO_{+1} \equi E_{46}-E_{28},   \qquad \cO_{+2}\equi E_{47}-E_{38}, \qquad \cO_{-1}\equi  E_{64}-E_{82},\qquad \cO_{-2}\equi E_{74}-E_{83}.
\end{equation}

\item[•]\underline{The $\cX$-generators:}
\begin{equation}
\cX_{+}\equi E_{67}-E_{32},  \qquad \cX_{-}\equi E_{76}-E_{23}.
\end{equation}
\end{itemize} 

One can check that the coset matrix $\cM$ \eqref{eq:MatrixMDef} has been obtained from the following SO(4,4) matrix elements \cite{Galtsov:2008bmt,Galtsov:2008jjb}:
\begin{equation}
\cM \= \cV^{T} K \cV\,,
\end{equation}
where
\begin{equation}
\begin{split}
K \equi &\text{diag}\left(1,1,-1,-1,1,1,-1,-1 \right),\\
\cV \equi & e^{\frac{1}{2}\left(\log(Z_0 Z_3) (\cH_1-\cH_2)+ \log(\cI_4) (\cH_2-\cH_3)+\log(Z_0^2 Z_1 Z_2) \cH_3 + \log\left(\frac{Z_2}{Z_1}\right) \cH_4  \right)} \, e^{-\frac{\mu Z_0^2}{\cI_4} \cX_+} \\
&e^{\sum_{I=1}^3 A_t^I \cZ_{+I}+ (A_\psi^I + \mu A_t^I) \cW_{+I}}\, e^{\Omega_\psi \cO_{+1}+\Omega_t \cO_{+2}}\, e^{\sum_{I=1}^3 B_I \cP_{+I}}
\end{split}
\label{eq:V&KMatrixDef}
\end{equation}

\subsubsection{Transformations preserving the flat asymptotics}
\label{sec:SO44Transfo}

In general, the transformation \eqref{eq:TNTransfo} by an arbitrary group element $g\in SO(4,4)$ does not preserve the asymptotic of the seed solution characterized by the asymptotic matrix $\cM_{0\infty} \= \underset{r\to\infty}{\text{lim}} \cM_0$ where $r$ is a radial coordinate. However, depending on $\cM_{0\infty}$ there always exist a  subgroup  $H\subset SO(4,4)$ that preserves $\cM_{0\infty}$. Thus,  the action on $\cM_0$ by an element of $H$ will generate a new solution, $\cM$, with the same asymptotic structure as that of $\cM_0$. 

Below,  we provide a list the generators of $SO(4,4)$ that preserves four-dimensional or five-dimensional asymptotic flatness \cite{Galtsov:2008bmt,Galtsov:2008jjb}.  Since five-dimensional flat space admits various forms where the $\psi$-fiber has different roles,  we will differentiate the five-dimensional case according to these different forms.

\begin{itemize}
\item[•] \textbf{A subgroup preserving} $\IR^{1,3}\times$S$^1$\textbf{:}

For solutions asymptotic to $\IR^{1,3}\times$S$^1$,\footnote{The addition of potential NUT or magnetic charges does not affect the discussion.}
\begin{equation}
ds_5^2 \to -dt^2 + d\psi^2 + dr^2 + r^2 (d\theta^2 +\sin^2 \theta\,d\phi^2)\,,\qquad X^I \to 1\,,
\label{eq:AsympR13S1}
\end{equation}
one has $\cM_{0\infty} \= K$ where $K$ is the involution matrix \eqref{eq:V&KMatrixDef}. 
We have identified a twelve-dimensional subgroup of $\mathfrak{so}(4,4)$ that leaves $K$ invariant. They are the sums of the $12$ positive and negative roots:
\begin{equation}
\cP_{+I}+ \cP_{-I}, \quad  \cZ_{+I}+\cZ_{-I},\quad \cW_{+I}-\cW_{-I}, \quad \cO_{+1}+\cO_{-1}, \quad \cO_{+2}-\cO_{-2}, \quad  \cX_++\cX_-,
\end{equation}
where $I=1,2,3$ and $p=1,2$.  The subgroup can be divided into three groups according to their commuting relations:
\begin{itemize}
\item[-] \underline{$\cP$-group:} since the generators $\cP_{\pm I}$ and $\cX_\pm$ commute,  they generate a subgroup of SO(4,4) characterized by the matrices:
\begin{equation}
g_\cP \= \exp \left[ \sum_{I=1}^3 \delta_I \left( \cP_{+I}+ \cP_{-I}\right) + \delta_4 \left(\cX_++\cX_- \right)\right],
\label{eq:Pgroup}
\end{equation}
where $(\delta_1,\delta_2,\delta_3,\delta_4)$ are four arbitrary constants.
\item[-] \underline{$\cZ$-group:} since the generators $\cZ_{\pm I}$ and $\cO_{\pm 1}$ commute,  they generate a subgroup of SO(4,4) characterized by the matrices:
\begin{equation}
g_\cZ \= \exp \left[ \sum_{I=1}^3 \gamma_I \left( \cZ_{+I}+ \cZ_{-I}\right) + \gamma_4 \left(\cO_{+1}+\cO_{-1} \right)\right],
\label{eq:Zgroup}
\end{equation}
where $(\gamma_1,\gamma_2,\gamma_3,\gamma_4)$ are four arbitrary constants.
\item[-] \underline{$\cW$-group:} since the generators $\cW_{\pm I}$ and $\cO_{\pm 2}$ commute,  they generate a subgroup of SO(4,4) characterized by the matrices:
\begin{equation}
g_\cW \= \exp \left[ \sum_{I=1}^3 \alpha_I \left( \cW_{+I}-  \cW_{-I}\right) + \alpha_4 \left(\cO_{+2}-\cO_{-2} \right)\right],
\label{eq:Wgroup}
\end{equation}
where $(\alpha_1,\alpha_2,\alpha_3,\alpha_4)$ are four arbitrary constants.
\end{itemize}

\item[•] \textbf{Subgroups preserving $\IR^{1,4}$:} 

The $\IR^{1,4}$ admit two different $\IR^t \times$S$^1\times\IR^3$ decompositions which have their own SO(4,4) invariant subgroups.

\begin{itemize}
\item[-] \underline{Diagonal fibration:}
\end{itemize}

We consider solutions that are asymptotic to
\begin{equation}
ds_5^2 \to -dt^2+dr^2+r^2\left(d\theta^2+\sin^2\theta\, d\phi^2+\begin{Bmatrix}
\sin^2\theta \sin^2 \phi\\
\cos^2\theta
\end{Bmatrix}  d\psi^2\right),\qquad X^{I} \to 1,
\end{equation}
where for the upper term in the bracket,  the $(\theta,\psi,\phi)$ angles define the hyperspherical coordinates of the S$^3$ while for the lower term, they form the Hopf coordinates.  The corresponding $\cM_{0\infty}$ is given by \eqref{eq:MatrixMDef} with $Z_0 \= (r \sin \theta \sin \phi)^{-2}$ or $(r \cos \theta)^{-2}$ while all other fields are trivial.  We have identified a three-dimensional subgroup of $\mathfrak{so}(4,4)$ that preserves $\cM_{0\infty}$:
\begin{equation}
\cZ_{+I}+\cZ_{-I}.
\end{equation}
Thus,  the corresponding SO(4,4) transformations are a subgroup of the $\cZ$-group \eqref{eq:Zgroup} with $\gamma_4=0$.
\begin{itemize}
\item[-] \underline{Hopf fibration:}
\end{itemize}

The $\IR^{4}$ space can be also written as a S$^1$ Hopf fibration over $\IR^3$ given by
\begin{equation}
ds_5^2 \to -dt^2+ r (d\psi+\cos\theta\,d\phi)^2+\frac{1}{r}\left[dr^2+r^2(d\theta^2+\sin^2\theta d\phi^2)\right],\qquad X^{I} \to 1,
\end{equation}
The corresponding $\cM_{0\infty}$ is given by \eqref{eq:MatrixMDef} with $Z_0 \= r^{-1}$, $\omega_\psi = \cos \theta \, d\phi$ for which the scalar dual is $\Omega_\psi = -r$ \eqref{eq:ScalarDuals} and all other fields are trivial. 

We found a seven-dimensional subgroup of $\mathfrak{so}(4,4)$ that preserves $\cM_{0\infty}$  generated by
\begin{equation}
\cZ_{+I}+\cZ_{-I},  \qquad \cW_{-I} - \cP_{-I} , \qquad \cX_+ + \cO_{-2}\,.
\end{equation}
Considering commutation relations,  the corresponding SO(4,4) transformations are the subgroup of the $\cZ$-group with $\gamma_4=0$ as before,  and what we define as the mix-group:
\begin{equation}
g_\text{mix} \= \exp \left[ \sum_{I=1}^3 \kappa_I \left( \cW_{-I}-  \cP_{-I}\right) + \kappa_4 \left(\cX_++\cO_{-2} \right)\right].
\label{eq:mixgroup}
\end{equation}

\end{itemize}

\subsection{Ernst transformations}

In this section,  we establish the action of the SO(4,4) symmetry group in terms of the Ernst formalism and Ernst potential \eqref{eq:ErnstPotDef}.  In four-dimensional general relativity,   solutions with a timelike Killing vector are defined by a complex-valued scalar Ernst potential $\cE = -g_{tt} + i \,\Omega_t$,  where $g_{tt}$ is the metric component along the time direction and $\Omega_t$ is the scalar dual of the one-form defining the timelike fibration.  The vacuum Ernst equations admit an SL(2,$\IR$) symmetry group that allow to generate solutions from known seeds.  The transformations can be decomposed into three groups:
\begin{itemize}
\item[•] \underline{Gauge transformation:} $\qquad \cE \to \cE + i \alpha,\qquad \alpha \in \IR.$
\item[•] \underline{Ehlers transformation:} $\qquad \cE \to \dfrac{\cE + i \alpha}{1+i \alpha \cE},\qquad \alpha \in \IR.$
\item[•] \underline{Scale transformation:} $\qquad \cE \to \alpha \cE,\qquad \alpha \in \IR.$
\end{itemize}

Our $4\times 4$ matrix Ernst formulation applied to five-dimensional supergravity,  introduced in Section \ref{sec:ErnstFormalism},  offers a generalization of the vacuum Ernst formalism.  It admits a similar group structure and transformations but with SL(2,$\IR$) replaced by SO(4,4) as highlighted in the previous sections.  Moreover,  because the Ernst potential is now real-valued,  the complex conjugation,  imaginary or real parts are replaced by matrix transpose,  symmetric or skew-symmetric parts respectively.  

The Ernst transformations can also be decomposed into three groups,  gauge,  Ehlers, and scale,  that we will connect with the SO(4,4) transformations introduced previously:
\begin{itemize}
\item[•] \underline{Gauge transformation:} $\qquad \cE \to \cE + G,\qquad G^T = - G.$

In terms of symmetric or skew-symmetric parts \eqref{eq:ErnstPotDef},  this corresponds to $P\to P$ and $Q \to Q-G$.  This subgroup is six-dimensional and we can show that it is equivalent to the SO(4,4) transformations on $\cM$ generated by the six $\mathfrak{so}(4,4)$ generators,  $\mathfrak{g}_i=(\cP_{+1},\cP_{+3},\cW_{+2}, \cZ_{+2},\cO_{+1},\cO_{+2})$,  since we have
\begin{equation}
e^{\sum_{i=1}^6 c_i \mathfrak{g}_i} \= \begin{pmatrix} 
\mathbb{I}_4 && -G \\
0 && \mathbb{I}_4
\end{pmatrix},\qquad G^T = - G,
\end{equation}
where $G$ is the skew symmetric matrix of the gauge transformation.  Moreover,  the one-form matrices $\cN_I$ \eqref{eq:NblockMatrixDef} transformed under gauge transformations:
\begin{equation}
\cN_1 \to \cN_1 \,,\qquad \cN_2 \to \cN_2 - G \cN_1\,,\qquad \cN_3 \to \cN_3 - \cN_2 G - G \cN_2^T+G \cN_1 G\,.
\end{equation}
\item[•] \underline{Ehlers transformation:} $\qquad \cE \to \left(1+\cE G\right)^{-1} \cE,\qquad G^T = - G.$

This six-dimensional subgroup has a non-trivial effect on the matrices $P$ and $Q$.  One can show that it is equivalent to the SO(4,4) transformations generated by the six $\mathfrak{so}(4,4)$ generators,  $\mathfrak{g}_i=(\cP_{-1},\cP_{-3},\cW_{-2}, \cZ_{-2},\cO_{-1},\cO_{-2})$,  as
\begin{equation}
e^{\sum_{i=1}^6 c_i \mathfrak{g}_i} \= \begin{pmatrix} 
\mathbb{I}_4 && 0 \\
-G && \mathbb{I}_4
\end{pmatrix},\qquad G^T = - G,
\end{equation}
where $G$ is the skew symmetric matrix of the Ehlers transformation.  The transformation rules of the one-form potentials \eqref{eq:NblockMatrixDef} under Ehlers transformations are:
\begin{equation}
\cN_1 \to \cN_1 - G \cN_2  -  \cN_2^T G+G \cN_3 G \,,\qquad \cN_2 \to \cN_2 - \cN_3 G \,,\qquad \cN_3 \to \cN_3\,.
\end{equation}
\item[•] \underline{Scale transformation:} $\qquad \cE \to S^T \cE S,\qquad \det S \neq 0.$

This 16-dimensional subgroup acts on $P$ and $Q$ such as $P \to S^{-1} P (S^T)^{-1}$ and $Q\to S^T Q S$.  We can show that it is equivalent to the SO(4,4) transformations generated by the sixteen remaining $\mathfrak{so}(4,4)$ generators such as
\begin{equation}
e^{\sum_{i=1}^{16} c_i \mathfrak{g}_i} \= \begin{pmatrix} 
(S^T)^{-1} && 0 \\
0 && S
\end{pmatrix},\qquad \det S \neq 0, 
\end{equation}
where $S$ is the invertible matrix of the scale transformation.  The transformation rules of the one-form potentials \eqref{eq:NblockMatrixDef} under scale transformations are:
\begin{equation}
\cN_1 \to S^{-1} \cN_1 (S^T)^{-1} \,,\qquad \cN_2 \to S^T \cN_2 (S^T)^{-1} \,,\qquad \cN_3 \to S^T \cN_3 S\,.
\end{equation}
\end{itemize}

Thus,  we have shown that the Ernst transformations on the matrix Ernst potential $\cE$ provide another formulation of the SO(4,4) transformations on the coset matrix $\cM$.

In general,  the Ernst transformations do not preserve the asymptotic of the seed solutions.  The transformations that preserve the asymptotic Ernst potential, $\cE_{\infty} \equi \underset{r\to\infty}{\lim} \cE$,  are 
\begin{itemize}
\item[•] \underline{Normalized scale transformation:} 
\begin{equation}
 \cE \to S^T \cE S,\qquad \text{for}\quad S^T \cE_{\infty} S \= \cE_{\infty}. \label{eq:NormalizedScale}
\end{equation}
\item[•] \underline{Normalized Ehlers transformation:} 
\begin{equation}
\cE \to (1+\cE_\infty G) (1+\cE G)^{-1} \cE (1-G \cE_\infty^T) + \cE_\infty G \cE_\infty^T,\qquad G^T=-G.
\label{eq:NormalizedErnst}
\end{equation}
This transformation consists  of an Ehlers transformation followed by a gauge transformation of matrix shift $(1+\cE_\infty G)^{-1}\cE_\infty G \cE_\infty^T (1- G\cE_\infty^T)^{-1}$ and a scale transformation of scaling factor $1-G \cE_\infty^T$.
\end{itemize}

While the normalized Ehlers transformations correspond to a six-dimensional subgroup defined by an arbitrary skew symmetric matrix $G$,  the dimension of the normalized scale transformations depends on the structure of $\cE_{\infty}$.  For instance,  when the asymptotic is $\IR^{1,3}\times$S$^1$ \eqref{eq:AsympR13S1},  i.e.  $\cE_{\infty}=\text{diag}(1,1,-1,-1)$,  the normalized scale transformations are given by the scale matrix $S\in\,$SO(2,2) of dimension $6$. 

This shows that the SO(4,4) subgroups identified in the previous section are not necessarily the largest subgroups preserving the different asymptotics considered.  While these largest subgroups admit a very simple formulation in terms of normalized Ernst transformations,  they can have a more involved expression in terms of $\mathfrak{so}(4,4)$ generators that we could not identify.  This shows that the Ernst formulation of the transformations might be a more appropriate formulation of the SO(4,4) symmetry of the sigma model.

\subsection{Inverse scattering}

Belinski and Zakharov developed an algebraic inverse-scattering method, also known as the \emph{$n$-soliton transformation}, for constructing cohomogeneity-two solutions of $D$-dimensional vacuum Einstein gravity \cite{Belinsky:1971nt,Belinsky:1979mh,Belinski:2001ph}. This method relies on the metric $G$ along the $D-2$ Killing directions, which satisfies equations of the form \eqref{eq:2dEquations}. The same approach can be applied here by replacing $G$ with the coset matrix $\cM$ of five-dimensional supergravity. However, an additional complication arises: unlike the metric $G$, $\cM$ must belong to SO(4,4), imposing extra constraints. The extension of the inverse-scattering method to STU supergravity has been thoroughly developed in \cite{Figueras:2009mc,Katsimpouri:2013wka}. Since this is not the primary focus of our paper, we provide only a brief review of the construction and adapt it to our conventions.

We consider cohomogeneity-two solutions, as introduced in Section \ref{sec:Coho2}, where the fields depend on two spacetime variables, $\rho$ and $z$.
The inverse-scattering method is based on the ``Lax pair" function $\Psi$, which depends on $(\rho,z)$ and a spectral parameter $\lambda$ and satisfies the equations 
\begin{equation}
D_1 \Psi \= \frac{\rho V - \lambda U}{\lambda^2+\rho^2} \Psi ,\qquad D_2 \Psi \= \frac{\rho U + \lambda V}{\lambda^2+\rho^2} \Psi , 
\end{equation}
 where\footnote{The integrability condition for these equations is ensured by $\partial_\rho U + \partial_z V = 0$, which follows from \eqref{eq:2dEquations}.}
\begin{equation}
U \= \rho (\partial_\rho \cM) \cM^{-1}\,,\quad V \= \rho (\partial_z \cM) \cM^{-1}\,,\quad D_1 \= \partial_z - \frac{2\lambda^2}{\lambda^2+\rho^2} \partial_\lambda\,,\quad D_2 \= \partial_\rho + \frac{2\lambda \rho}{\lambda^2+\rho^2} \partial_\lambda\,.
\end{equation}
The matrix $\cM$ can be extracted from $\Psi$ as
\begin{equation}
\cM \= \Psi |_{\lambda=0}\,.
\label{eq:transformedM}
\end{equation}
A $n$-soliton transformation consists of the following steps:
\begin{itemize}
\item[•] Start with a known seed solution $\cM_0$ and derive its Lax pair function $\Psi_0$ from the equations above.
\item[•] Obtain a new $\Psi$ by dressing $\Psi_0$ with a matrix that has $n$ simple poles in $\lambda$: $\chi \equi \mathbb{I} + \sum_{k=1}^n \frac{R_k}{\lambda-\mu_k}$: $\Psi = \chi \,\Psi_0$.  
The equations for $\Psi$ determine the locations of the poles $\mu_k$ and the matrices $R_k$. The poles are given by
\begin{equation}
\mu_k \= a_k -z \pm \sqrt{\rho^2+(z-a_k)^2}\,,
\end{equation}
where the ``$+$" and ``$-$'' poles are commonly referred to as \emph{solitons} and \emph{antisolitons} respectively,  and $a_k$ are real constants that will correspond to the coordinates of the sources on the $\rho=0$ axis.  The matrices $R_k$ are defined in terms of $n$ arbitrary eight-dimensional vectors $m^{(k)}$ as
\begin{equation}
 R_k \= \sum_{l=1}^n \frac{(\Gamma^{-1})_{lk}}{\mu_l}\, \,\cM_0 \,p^{(l)}{p^{(k)}}^T\,,
\end{equation}
where we have introduced
\begin{equation}
p^{(k)} \= \Psi_0^{-1} |_{\lambda=\mu_k} \,m^{(k)}\,,\qquad \Gamma_{lk} \= \frac{{p^{(l)}}^T \cM_0 \,p^{(k)}}{\rho^2+\mu_k \mu_l}\,. \label{eq:GammaMat}
\end{equation}
\item[•] Compute the transformed $\cM$ from \eqref{eq:transformedM}:
\begin{equation}
\cM\= \cM_0 \- \sum_{l,k=1}^n \frac{(\Gamma^{-1})_{lk}}{\mu_l \mu_k}\,\, \left(\cM_0 \,p^{(l)}\right)\left(\cM_0 \,p^{(k)}\right)^T\,.
\end{equation}
\item[•] Compute the three-dimensional base metric $ds_3^2$ as given in \eqref{eq:3DBase}. Unlike Ernst or SO(4,4) transformations, the three-dimensional base is not preserved. Instead, it is determined by the conformal factor $e^{2\nu}$, given by \cite{Belinsky:1971nt,Belinsky:1979mh}:
\begin{equation}
e^{2\nu} \= C\, e^{2\nu_0}\times \prod_{k=1}^n \frac{\mu_k^2}{\mu_k^2+\rho^2}\,\det \Gamma\,,
\label{eq:BaseFactorNSoliton}
\end{equation}
where $C$ is an arbitrary constant and $e^{2\nu_0}$ is the conformal factor of the seed solution.

\item[•] Impose the SO(4,4) structure on $\cM$.

While the transformed matrix $\cM$ is symmetric, it does not generically satisfy the SO(4,4) group structure including $\det \cM =1$.\footnote{More precisely, one has  $\det \cM = (-1)^n \rho^{2n} \prod_{k=1}^n \mu_k^{-2}\,\,\det \cM_0$.} In the original Belinski-Zakharov construction,  having $\det \cM =\det \cM_0$ is obtained by multiplying $\cM$ by suitable factors of $\rho$ and $\mu_k$, which do not affect the equations of motion. However, for matrices of dimension greater than 2, this often leads to singularities at $\rho=0$.

A more refined approach, introduced in \cite{Pomeransky:2005sj}, circumvents this issue. One starts with a diagonal seed solution $\cM_0$ (such as flat space, multi-Schwarzschild geometry, or multi-bubble geometry) and first removes some solitons or antisolitons with trivial vectors $m^{(k)}_b = \delta_{ab}$ for a given index $a$. This effectively rescales the $(aa)$ component of $\cM_0$ by a factor $-\mu_k^2/\rho^2$ while leaving other components unchanged. The same soliton or antisoliton is then reintroduced with a general $m^{(k)}$ vector. This procedure ensures that $\det \cM = \det \cM_0 = 1$ while avoiding singular behavior at $\rho=0$.

After these steps, the transformed base warp factor \eqref{eq:BaseFactorNSoliton} simplifies to
\begin{equation}
e^{2\nu} \= e^{2\nu_0} \,\frac{\det \Gamma}{\det \Gamma_0},
\end{equation}
where $\Gamma_0$ and $\Gamma$ are constructed as in \eqref{eq:GammaMat} using $\cM_0$ and $\cM$ respectively.

Finally, ensuring that $\cM$ remains in the SO(4,4) group, as required by \eqref{eq:SO(4,4)M}, imposes additional constraints on the vectors $m^{(k)}$. Unfortunately, these constraints are nonlinear and cannot be solved in a general manner but must be addressed case by case. Since this is beyond the scope of our discussion, we refer the reader to \cite{Figueras:2009mc,Katsimpouri:2013wka} for a more in-depth analysis. 
\end{itemize}

The main advantage of the $n$-soliton transformation compared to the Ernst and SO(4,4) transformations is that it modifies the source structure of the seed solution, enabling transitions from simple solutions to ones with multiple sources that may carry charges and spin.

\section{Static microstates of non-extremal black holes}
\label{sec:NewSol}

In this section, we illustrate the effectiveness of the simplest solution-generating techniques discussed in the previous sections by constructing novel non-BPS black hole microstates in supergravity.  The SO(4,4) transformations have been widely employed to generate non-supersymmetric black holes, black strings, and black rings in supergravity  \cite{Cvetic:1995kv,Chong:2004na,Giusto:2007fx,Bouchareb:2007ax,Tomizawa:2008qr,Galtsov:2008bmt,Galtsov:2008jjb,Virmani:2012kw,Chow:2014cca}. The principle behind these constructions is straightforward: one starts with a known vacuum black hole solution and generates supergravity charges by applying appropriate SO(4,4) transformations. These transformations typically preserve the internal topology of the seed solution while introducing the desired electromagnetic fluxes.

Remarkably, while these methods have been predominantly applied to black holes, there exists a rich variety of known smooth horizonless geometries — both BPS and non-BPS — that can serve as seed solutions. These geometries can be used to generate interesting smooth solutions with the same charges as non-extremal black holes, thereby providing coherent microstates of non-extremal black holes.

In this section, we will explore the potential of this approach by focusing on relatively simple geometries. Specifically, we will construct solutions that feature a single bubbling source, are static, and are asymptotically $\mathbb{R}^{1,3} \times$S$^1$. To achieve this, we will use the Taub-bolt solution \cite{Page:1978hdy} as the seed solution and apply previously identified SO(4,4) transformations to generate charged smooth solutions comparable to non-extremal black holes. This process will allow us to recover the topological star solution constructed by one of the authors \cite{Bah:2020ogh,Bah:2020pdz}, as well as generate new static smooth solutions with more intricate electromagnetic field configurations.

\subsection{Seed solution: Taub-bolt geometry}

The seed solution is a trivial timelike fibration over the Taub-bolt instanton \cite{Page:1978hdy}, which corresponds to the Euclidean Schwarzschild-NUT geometry. This geometry represents a vacuum solution in five dimensions with the following metric:
\begin{equation}
ds_5^2 \= - dt^2 + \frac{r^2-2m r+n^2}{r^2-n^2} \left(d\psi+2n\cos\theta\,d\phi\right)^2 + \frac{r^2-n^2}{r^2-2m r+n^2} \,ds_3^2 \,,
\end{equation}
where the solution is expressed as a timelike and spacelike fibration over the three-dimensional base space: 
\begin{equation}\label{eq:3dBase}
ds_3^2 \= dr^2 +(r^2-2m r+n^2) \left(d\theta^2 + \sin^2 \theta\,d\phi^2 \right)\,.
\end{equation}
The three-dimensional base is flat if and only if $m=|n|$, and the validity of the solution requires 
\begin{equation}
m \,\geq\, |n| .
\label{eq:CondValTNUT}
\end{equation}
In terms of the five-dimensional ansatz \eqref{eq:Metric&FieldsAnsatz}, the Taub-bolt solution is described by the fields:
\begin{equation}
Z_I \= 1 \,,\quad Z_0 \= \frac{r^2-n^2}{r^2-2m r+n^2}\,,\quad \mu \= A_t^I \= A_\psi^I \= a^I\= \omega_t \= 0\,,\quad \omega_\psi \= 2 n \cos\theta \, d\phi.
\end{equation}
The scalar duals of the one-forms, given by \eqref{eq:ScalarDuals}, are:
\begin{equation}
\Omega_t \= B_I \= 0\,,\qquad \Omega_\psi \= \frac{2n(r-m)}{r^2-n^2}.
\end{equation}
From these, the coset matrix $\cM_0$ \eqref{eq:MatrixMDef} and the Ernst potential $\cE_0$ \eqref{eq:ErnstPotDef} are determined using the following $P$ and $Q$ matrices:
\begin{equation}
P \= \text{diag}\left[1,Z_0,-1,-Z_0 \right]\,,\qquad Q\= \Omega_\psi \left(E_{42}-E_{24} \right),
\end{equation}
where $E_{ij}$ is here the $4\times4$ matrix with a $1$ in the $(i,j)$ component and zeros elsewhere.

Additionally, the one-form matrix $\cN_{0}$ \eqref{eq:NMatrixDef} is derived and decomposed into four $4\times 4$ blocks as defined in \eqref{eq:NMatrixDecomp}:
\begin{equation}
{\cN_{2}}_0 \= 2m\cos \theta\,d\phi\,\text{diag}\left[ 0,1,0,1 \right]\,,\qquad {\cN_{1}}_0 \= -{\cN_{3}}_0 \= -2n\cos\theta\,d\phi \left( E_{42}-E_{24}\right).
\end{equation}
From these expressions, it is clear that the Taub-bolt geometry is sourced by two monopoles: one with charge $m$, which generates the gravitational field, and another with charge $n$, which sources the Kaluza-Klein vector field.  Using $\text{SO}(4,4)$ transformations on $\cM_0$,  \eqref{eq:TNTransfo},  or,  equivalently,  normalized Ernst and scale transformations on $\cE_0$,  \eqref{eq:NormalizedErnst}, and \eqref{eq:NormalizedScale}, one can redistribute these monopoles into other fields of five-dimensional supergravity.

The transformations do not affect the local divergences of the fields so that the transformed fields will still admit a singularity at the Taub-bolt locus $r=r_+$:
\begin{equation}
r_+ \equi m \+ \sqrt{m^2-n^2}\,.
\label{eq:CoorDegen}
\end{equation}

\subsection{New solitons from group transformations}

There are multiple ways to transform the Taub-bolt solution into new \emph{gravitational solitons} with a broader charge content and, ultimately, the same mass and charges as nonextremal black holes.  However, the transformed solution must satisfy the following key conditions to correspond to a smooth horizonless geometry asymptotic to $\IR^{1,3}\times$S$^1$:
\begin{itemize}
\item[•] The singularity at $r=r_+$ must correspond to a smooth coordinate degeneracy of the $\psi$ circle.  The SO(4,4) transformations do not alter the singular features of the fields, so if we start with a Taub-bolt spacetime where $Z_0$ is singular and $Z_I$ is regular at $r=r_+$, the transformed fields will retain these properties.  Thus,  from \eqref{eq:Metric&FieldsAnsatz}, one must therefore impose $\mu = 0$ at $r=r_+$ to ensure that the spacetime terminates smoothly at $r=r_+$.  
\item[•] The final solution must have no NUT charge along the time direction, $\omega_t \underset{r\to \infty}{\to} 0$ \eqref{eq:Metric&FieldsAnsatz}. 
\end{itemize}
To summarize,  having a regular asymptotically $\IR^{1,3}\times$S$^1$ soliton requires two key conditions:
\begin{equation}
\text{(C1):} \quad \mu \underset{r\to r_+}{\to} 0\,,\qquad \text{(C2):} \quad \omega_t \underset{r\to \infty}{\to} 0\,.
\label{eq:RegCondGen}
\end{equation}

Since the SO(4,4) transformations preserving the $\IR^{1,3}\times S^1$ asymptotics have been decomposed into three subgroups in Section \ref{sec:SO44Transfo},  namely, the $\cP$, $\cZ$, and $\cW$ groups,  it is natural to categorize the new solitons according to these subgroups:
\begin{itemize}
\item[•] Applying each subgroup individually, while ensuring the above conditions, generates three distinct solutions: the $\cP$-soliton, the $\cZ$-soliton, and the $\cW$-soliton. These geometries will be characterized by five parameters at most, $m$ and $n$, along with three transformation parameters.
\item[•] Applying two subgroups together produces more general solutions, such as the $\cP\cW$-soliton and other possible combinations. Since the transformations of different subgroups do not commute, the $\cP\cW$-soliton and the $\cW\cP$-soliton will yield distinct solutions. These configurations are expected to be parametrized by nine quantities: $m$, $n$, and seven transformation parameters.
\item[•] The most general solutions arise from applying all three subgroups, leading to configurations such as the $\cP\cW\cZ$-soliton and its permutations. These solutions should be determined by 13 parameters in total, corresponding to eight arbitrary electromagnetic charges and four arbitrary scalar charges.
\end{itemize}

Naturally, as more transformations are applied, the resulting solutions become increasingly complex. In this paper, we focus on the first class of solutions, where only a single subgroup is applied. For black holes \cite{Chow:2014cca},  only the $\cP$ and $\cZ$ groups have been used since the $\cW$ group does not affect the vacuum black seed solutions considered.  For smooth solutions,  we found that it is the $\cZ$ group that does not affect the Taub-bolt seed solution while the $\cW$ group acts non-trivially.\footnote{This does not imply that the $\mathcal{Z}$-group acts trivially on all solutions and should be disregarded. For instance, the $\mathcal{Z}$-group has a nontrivial effect for the $\mathcal{Z}\mathcal{P}$-solitons, since the $\cP$ and $\cZ$ transformations do not commute.}

Thus, we consider two specific solutions: the $\cW$-soliton and the $\cP$-soliton. All resulting geometries will be static,  smooth, and horizonless, with electric and magnetic charges, parametrized by five independent quantities.

\subsection{The $\cW$-soliton}

We apply the $\mathrm{SO}(4,4)$ transformation \eqref{eq:TNTransfo} to the Taub-bolt seed  with $g = g_\cW$ \eqref{eq:Wgroup}. Extracting the fields using \eqref{eq:ExtractingFields}, we find that the transformed solution satisfies the regularity condition \eqref{eq:RegCondGen} when $\alpha_4 = 0$. The resulting solution is given by
\begin{align}
ds_5^2 &\=  \frac{\Delta}{(H_1 H_2 H_3)^\frac{2}{3}} \left(d\psi+P_0\cos \theta\, d\phi + \chi\, dt\right)^2+\frac{(H_1 H_2 H_3)^\frac{1}{3}}{\sqrt{\Delta}}\left[-\sqrt{\Delta}\,dt^2 + \frac{1}{\sqrt{\Delta}}\, ds_3^2 \right],\nn\\
X^I &\= \frac{(H_1 H_2 H_3)^\frac{1}{3}}{H_I}\,,\qquad A^I \=A_\psi^I \, \left(d\psi+P_0\cos \theta\,  d\phi + \chi dt\right)+A_t^I dt -P_I \cos\theta \,d\phi\,, \label{eq:WStar}
\end{align}
where $ds_3^2$ is the invariant base metric \eqref{eq:3dBase}, and we have defined
\begin{align}
\Delta & \= 1-\frac{2(mr-n^2)}{r^2-n^2}\,,\quad H_I \= 1-\frac{2(mr-n^2)\sin^2\alpha_I}{r^2-n^2}\,,\quad \chi \= \frac{2n(r-m)\prod_{I=1}^3 \sin \alpha_I}{r^2-n^2}\,,\nn\\
 A_\psi^I &\= \frac{(mr-n^2)\sin 2\alpha_I}{(r^2-n^2)H_I}\,,\quad A_t^I \= \frac{2n(r-m)}{r^2-n^2}\frac{|\epsilon_{IJK}|}{2} \cos \alpha_I \sin \alpha_J \sin \alpha_K\,.\label{eq:WTSFields}
\end{align}
The solution carries four magnetic charges $P_\Lambda$ and four electric charges $Q_\Lambda$, given by \eqref{eq:conservedchargesGen}
\begin{equation}
\begin{split}
P_0 &\= 2n\prod_{I=1}^3 \cos \alpha_I\,,\qquad P_I \= 2n \frac{|\epsilon_{IJK}|}{2} \sin \alpha_I \cos \alpha_J \cos \alpha_K\,,  \\
Q_0 &\= 2n \prod_{I=1}^3 \sin \alpha_I \,,\qquad Q_I \= -2n\frac{|\epsilon_{IJK}|}{2} \cos \alpha_I \sin \alpha_J \sin \alpha_K\,.
\end{split}
\label{eq:ChargesWTS}
\end{equation}
When embedded in M-theory on T$^6\times$S$^1$ as given in \eqref{eq:MTheoryEmbed},  $P_0$ corresponds to a KKm charge,  $P_I$ to M5-brane charges,  $Q_0$ to P-momentum charge, and $Q_I$ to M2-brane charges. 

Notably, this solution shares similarities with the $\tfrac{1}{2}$-BPS center that forms the building block of some supersymmetric black hole microstates, known as BPS bubbling geometries \cite{Bena:2007kg,Heidmann:2017cxt,Bena:2017fvm,Warner:2019jll}. Indeed, the electromagnetic fluxes are constructed on a fixed three-dimensional base, and the electric charges are fixed by the magnetic charges such as,
\begin{equation}
Q_I \= - \frac{|\epsilon_{IJK}|}{2}  \frac{P_J P_K}{P_0}\,,\qquad Q_0 \= \frac{P_1 P_2 P_3}{P_0^2}\,,
\label{eq:ChargesCond}
\end{equation}
which is imposed by smoothness for the $\tfrac{1}{2}$-BPS center.
However, unlike their BPS counterparts, the three-dimensional and four-dimensional base metrics are not Ricci-flat and the electromagnetic fluxes do not satisfy the BPS floating brane ansatz \cite{Bena:2004de,Goldstein:2008fq}, making the $\cW$-soliton a non-extremal generalization of the supersymmetric centers studied in \cite{Bena:2007kg,Heidmann:2017cxt,Bena:2017fvm,Warner:2019jll}. \\

One can use the relations \eqref{eq:ChargesCond} and \eqref{eq:ChargesWTS} to express the parameters $(n, \alpha_I)$ in terms of the more physical magnetic charges $(P_0, P_I)$. This yields:
\begin{equation}
\alpha_I \= \arctan \left(\frac{P_I}{P_0}\right) \,,\qquad n \= \pm \frac{1}{2} \sqrt{P_0^2+P_1^2+P_2^2+P_3^2+Q_0^2+Q_1^2+Q_2^2+Q_3^2}\,.
\end{equation}
The second expression explicitly shows that the KKm charge of the seed solution,  $n$, has been redistributed through SO(4,4) transformations into the eight possible charges in five-dimensional STU supergravity while preserving smoothness.

Another surprising property is that the four-dimensional metric obtained by reducing along $\psi$ has not been affected by the SO(4,4) transformations and does not depend on the transformation parameters $\alpha_I$.  It is given by the bracketed term in \eqref{eq:WStar}, which matches the Taub-bolt seed solution. Consequently, the ADM mass is
\begin{equation}
M_\text{ADM} \= \frac{m}{2G_4}\,.
\end{equation}
This invariance may be understood by proposing an interpretation of the $\cW$-group transformations in type IIB supergravity \eqref{eq:TypeIIBsol}. We suggest that the parameters $\alpha_I$ of the $\cW$-group,  appearing through $\sin$ and $\cos$,  could correspond to rotations mixing the $\psi$ direction with an internal direction of the T$^4 \times$ S$^1$, possibly accompanied by additional S- and T-dualities. Since such transformations leave the four-dimensional metric unchanged, they could account for the invariance we observed.

To conclude,  the $\cW$-solitons define a family of eight-charge solutions depending on five independent parameters,  the four magnetic charges $(P_0,P_1,P_2,P_3)$ and the ADM mass $M_\text{ADM}$.

\subsubsection{Regularity}

Due to the KKm charge,  the $\theta=0$ and $\theta=\pi$ coordinate singularity of the S$^2$ is regular if we consider the following lattice of periodicity \cite{Bena:2005ay},
\begin{equation}
(\psi,\phi) \,\= \,  (\psi+2\pi R_{\psi},\phi) \=  (\psi+\pi R_{\psi},\phi +2\pi) \,,
\label{eq:Periodicity}
\end{equation}
which is compatible with the asymptotic flatness,  and if the KKm charge is quantized such as:
\begin{equation}
P_0 \= \frac{1}{2} N_0 R_\psi\,,\qquad N_0 \in \mathbb{Z}.
\label{eq:KKmQuantization}
\end{equation}
The radius $R_\psi$ is the size of the extra dimension, giving a length scale to the solution.

The spacetime terminates at $r = r_+ = m+\sqrt{m^2-n^2}$ as the degeneracy of the $\psi$ circle, provided that
\begin{equation}
\sin^2 \alpha_I \,\neq\, 1\,.
\end{equation}
In the non-extremal regime, where $|n| < m$, the surface $r = r_+$ corresponds to an origin in $\IR^2/\mathbb{Z}_k$ if one imposes\footnote{The metric at $r=r_+$ is derived by introducing $\frac{\rho^2}{4}=r-r_+$ and $\rho \to 0$.  We further impose that the metric at constant $t$,  $\phi$ and $\theta$ is given by $d\rho^2 +\frac{\rho^2}{k^2 R_{\psi}^2} d\psi^2$. \label{ft1}}
\begin{equation}
k R_\psi \= 2 (m+\sqrt{m^2-n^2}) \cos \alpha_1 \cos \alpha_2 \cos \alpha_3\,.
\end{equation}
Remarkably,  unlike the topological stars of \cite{Bah:2020ogh,Bah:2020pdz} which will be reviewed in Section \ref{sec:TSfirst}, the $\cW$-soliton does not require a large orbifold factor $k \gg 1$ to achieve an ADM mass significantly larger than the extra dimension radius.  This can be illustrated by reinterpreting the regularity condition in terms of the quantized KKm charge,  which gives:
\begin{equation}
2k \= \frac{m+\sqrt{m^2-n^2}}{n}\,N_0\,,
\label{eq:RegCondQuantized}
\end{equation}
where the dependence in $R_\psi$ has dropped.  The condition above can be thought of as fixing $n$ in terms of the orbifold parameter $k$:
\begin{equation}
n \= \frac{N_0}{k} \left(1+\frac{N_0^2}{4k^2} \right)^{-1} \,m\,,\qquad 2k\geq N_0\,.
\end{equation}

\subsubsection{Extremal and BPS limit}

The extremal limit of the $\cW$-soliton will be discussed in more detail in Section \ref{sec:PStarExtr}, as they match the solitons obtainable via $\cP$-group transformations.

In brief, the extremal limit, $|n|=m$, causes the $\IR^2/\mathbb{Z}_k$ bolt structure at the coordinate singularity to degenerate into a $\IR^4/\mathbb{Z}_{|N_0|}$ NUT structure, where the entire S$^2$ degenerates along the $\psi$-direction. This limit fixes the ADM mass in terms of the magnetic charges as 
\begin{equation}
M_\text{ADM} \= \frac{1}{4G_4} \sqrt{P_0^2+P_1^2+P_2^2+P_3^2+Q_0^2+Q_1^2+Q_2^2+Q_3^2},
\label{eq:ExtrBoundWSol}
\end{equation}
 which resembles the extremal bound for dyonic black holes in STU supergravity, with an additional factor of $\tfrac{1}{2}$.  \\
 
 In Section \ref{sec:PStarExtr},  we show that the extremal solutions correspond to a well-defined coordinate transformation,  known as spectral flow \cite{Bena:2008wt},  of a BPS solution of five-dimensional supergravity.  The BPS solution is sourced by a single BPS center preserving 16 supercharges.  This center, referred to as a \emph{Gibbons-Hawking center},  is the fundamental building block of smooth multicenter geometries, which are microstates of BPS black holes in supergravity \cite{Bena:2007kg,Heidmann:2017cxt,Bena:2017fvm,Warner:2019jll}.

In conclusion, $\cW$-solitons with $|n|>m$ can be viewed as non-extremal and non-BPS generalizations of these BPS centers,  suggesting they could play an equally crucial role in generating families of microstates for non-supersymmetric and non-extremal black holes in supergravity.

\subsubsection{Microstates of non-extremal black holes}

In general,  determining whether an eight-charge smooth horizonless solution can have the same mass and charges as a nonextremal black hole — and  correspond to one of its microstates — can be challenging.  Indeed,  the extremality bound for an eight-charge four-dimensional black hole can take a rather intricate form \cite{Chow:2014cca}.

One can consider an illustrative example where all charges have equal absolute values:
\begin{equation}
\alpha_I \= \frac{\pi}{4}\,,\qquad P_0 \= P_I \= Q_0 \= -Q_I \=Q\equi \frac{n}{\sqrt{2}}\,.
\label{eq:WTSEqualCharges}
\end{equation}
The soliton then becomes a solution of five-dimensional minimal supergravity \eqref{eq:L5min}, where the scalars are trivial, $X^I=1$, and the gauge fields are identical, $A^I=A$. The metric and field simplify to
\begin{align}
ds^2 \=& \frac{(r^2-2Q^2)\left(1-\frac{4M}{r} +\frac{2Q^2}{r^2}\right)}{(r-2M)^2}\, \left( d\psi +Q \cos \theta \,d\phi+\frac{Q(r-2M)}{r^2-2Q^2} dt\right)^2 \nn\\
&+ \left(1-\frac{2M}{r} \right) \left[-\frac{r^2}{r^2-2Q^2} \,dt^2 + \frac{dr^2}{1-\frac{4M}{r} +\frac{2Q^2}{r^2}}+r^2 d\Omega_2^2 \right]\,, \label{eq:WSolSimple}\\
A \=&  \frac{Q}{r} dt +\frac{2(M r-Q^2)}{r(r-2M)} (d\psi+Q \cos \theta \,d\phi)-Q\cos \theta\, d\phi\,,\nn
\end{align}
where $M$ corresponds to the ADM mass, $G_4 M_\text{ADM} =M$ and $d\Omega_2^2$ is the line element of a round S$^2$,  $d\Omega_2^2 = d\theta^2+\sin^2 \theta \,d\phi^2$. In this parametrization, the smooth end of spacetime is located at $r=2\left(M+\sqrt{M^2-\tfrac{1}{2}Q^2}\right)$. By requiring that this occurs before the singularity at $r=\sqrt{2}|Q|$, we recover the extremal bound of the $\cW$-soliton \eqref{eq:ExtrBoundWSol}:
\begin{equation}
|Q| \leq \sqrt{2}\,M\,.
\end{equation}
At extremality, $|Q|=\sqrt{2}M$, the $\psi$-circle and the entire $S^2$ smoothly degenerate at the coordinate singularity, forming a Taub-NUT center and creating an origin in $\IR^{4}/\mathbb{Z}_{2|Q|/R_\psi}$. Interestingly, the neutral limit, $Q=0$, does not yield Euclidean Schwarzschild geometry as one might expect; instead, part of the gauge field remains, leading to an axion in four dimensions.

At first glance, this solution might appear to have the same conserved charges as a dyonic Reissner-Nordström black hole with equal charges.\footnote{The embedding of such a black hole in five-dimensional minimal supergravity \eqref{eq:L5min} is given by $A = - \frac{Q}{r} dt - P \cos \theta \,d\phi$ and
\begin{equation}
ds_\text{RN}^2 = \left( d\psi +Q \cos \theta \,d\phi-\frac{P}{r} dt\right)^2 + \left(1-\frac{2M}{r} +\frac{P^2+Q^2}{r^2} \right) dt^2 + \frac{dr^2}{1-\frac{2M}{r} +\frac{P^2+Q^2}{r^2}}+r^2 d\Omega_2^2\,,\nn
\end{equation}} However, due to differences in charge signs, the soliton does not have the same conserved charges.

To identify the corresponding black hole, we analyzed the general eight-charge nonextremal black holes derived in \cite{Chow:2014cca}, identified a limit where the charges match those in \eqref{eq:WTSEqualCharges}, and simplified the solution. This yields a new black hole solution in five-dimensional minimal supergravity:
\begin{align}
ds_\text{BH}^2 \=&\frac{r^2-\left(\sqrt{M^2+4Q^2}-M\right)^2}{r^2}\left( d\psi +Q \cos \theta \,d\phi+\frac{Q}{r+\sqrt{M^2+4Q^2}-M} dt\right)^2 \nn \\
&- \frac{r(r-2M)}{r^2-\left(\sqrt{M^2+4Q^2}-M\right)^2}\,dt^2 +\frac{r\,dr^2}{r-2M}+r^2 \left(1-\frac{\sqrt{M^2+4Q^2}-M}{r} \right)d\Omega_2^2\,,\nn \\
A \=&  \frac{Q}{r} dt +\frac{\sqrt{M^2+4Q^2}-M}{r}\left( d\psi +Q \cos \theta \,d\phi\right)- Q \cos \theta d\phi\,, \label{eq:EquivBHWSol}
\end{align}
where $M$ corresponds to the ADM mass, $G_4 M_\text{ADM} =M$, and $Q$ to the eight charges as in \eqref{eq:WTSEqualCharges}. This black hole has the same conserved charges as the $\cW$-soliton, with the correct signs. Furthermore, like the $\cW$-soliton, it possesses a nontrivial axion and dilaton when reduced to four dimensions, distinguishing it from the Reissner-Nordström black hole.

The coordinate singularity at $r=2M$ corresponds to a regular horizon if and only if it occurs before the other singular locus, imposing the extremal bound
\begin{equation}
|Q| \,<\, \sqrt{2} M\,.
\label{eq:BHBound}
\end{equation}
The vacuum limit $Q=0$ reproduces the trivial embedding of the Schwarzschild black hole. Remarkably, the black hole regime precisely matches the range of validity of the nonextremal $\cW$-soliton, indicating that for any nonextremal black hole, there exists a corresponding $\cW$-soliton with the same conserved charges, replacing the horizon with a smooth, horizonless structure.

More interestingly,  in the extremal limit of the black hole,  $|Q| = \sqrt{2} M$, the metric and field remains well-defined, and the horizon disappears, giving way to a smooth horizonless $\IR^{4}/\mathbb{Z}_{2|Q|/R_\psi}$ end to spacetime. More precisely, one can show that both the $\cW$-soliton and the black hole solutions match at their extremal limit, $|Q| = \sqrt{2} M$,  yielding the exact same solution.  This demonstrates that,  for $|Q| < \sqrt{2} M$,  they correspond to two branches of the same solution that merge at extremality:  one describing a black hole and the other a smooth, horizonless geometry.

This behavior is reminiscent of the topological star and its corresponding black hole derived in \cite{Bah:2020ogh,Bah:2020pdz}, where the common extremal solution is the microscopic extremal three-charge black hole. However, for the topological star, both branches could be described by a single metric and gauge field. While we believe a similar decomposition should exist here, we have not yet identified the appropriate parametrization.

Finally, the black hole entropy and temperature are given by
\begin{equation}
\begin{split}
S&\= \frac{\pi\left(\left(M+\sqrt{M^2+4Q^2}\right)^2 -8Q^2\right)}{G_4 Q}\sqrt{M\sqrt{M^2+4Q^2}-M^2-Q^2}\,,\\ 
T &\= \frac{1}{4\pi \sqrt{\left(M+\sqrt{M^2+4Q^2}\right)^2 -8Q^2}}\,.
\end{split}
\end{equation}
At $Q=0$, we recover the Schwarzschild values, while at extremality, the entropy vanishes ($S=0$), consistent with the transition to a smooth, horizonless geometry.

This example with equal charges suggests that the $\cW$-soliton exists in the same range of mass and charges as a nonextremal eight-charge black hole in STU supergravity,  and both correspond to two branches of the same solution at fixed mass and charges. As such, we propose that the $\cW$-soliton represents a class of coherent yet atypical microstates of nonextremal black holes.

\subsection{The $\cP$-soliton}

We now apply the $\mathrm{SO}(4,4)$ transformation \eqref{eq:TNTransfo} to the Taub-bolt seed with $g = g_\cP$ \eqref{eq:Pgroup}. After extracting the fields using \eqref{eq:ExtractingFields} and performing algebraic simplifications, we find that the solution satisfies the regularity constraints \eqref{eq:RegCondGen} if and only if:
\begin{equation}
(\text{S}_1):\,n \=\delta_4 \=  0 \qquad \textbf{or} \qquad (\text{S}_2):\,\delta_3 \= \delta_4 \= 0 \qquad \textbf{or} \qquad (\text{S}_3):\,|n| \= m \,,\, \tanh \delta_4 = \prod_{I=1}^3 \tanh \delta_I \,. 
\end{equation}
For the second solution, we could have chosen $\delta_1$ or $\delta_2$ instead of $\delta_3$, which is equivalent due to the permutation symmetry of the scalars and gauge fields.

We now analyze each of these three solutions separately. Note that while the third solution has more nonzero parameters, it is constrained to its extremal limit, $|n|=m$, where the three-dimensional base becomes flat $\IR^3$ \eqref{eq:3dBase}.

\subsubsection{(S$_1$): the generalization of the topological star}
\label{sec:TSfirst}

We apply the $\cP$-group transformations with $n=\delta_4 \= 0$.  The transformed solution is given by\footnote{We changed $\delta_I \to -\delta_I$ to have positive charges for positive $\delta_I$.}
\begin{align}
ds_5^2 &\=  \frac{1}{(H_1 H_2 H_3)^\frac{1}{3}} \left[-dt^2 + \left( 1-\frac{2m}{r} \right) d\psi^2 \right] \+ (H_1 H_2 H_3)^\frac{2}{3} \left[ \frac{dr^2}{1-\frac{2m}{r}} + r^2 \left(d\theta^2 + \sin^2 \theta\,d\phi^2 \right) \right],\nn\\
X^I &\= \frac{H_I}{(H_1 H_2 H_3)^\frac{1}{3}}\,,\qquad A^I \= -m \sinh 2\delta_I \,\cos \theta \,d\phi\,,\qquad H_I \= 1+\frac{2m \sinh^2 \delta_I}{r}\,.
\end{align}
The transformation has generated three magnetic charges,  $P^I$,  corresponding to M5-brane charges when embedded in eleven-dimensional supergravity \eqref{eq:MTheoryEmbed}.  The mass and charges are given by \eqref{eq:conservedchargesGen}
\begin{equation}
M_\text{ADM} \= \frac{m}{4G_4} \left(-1+\cosh 2\delta_1+\cosh 2\delta_2+\cosh 2\delta_3 \right)\,,\qquad P^I \= m \sinh 2\delta_I\,.
\end{equation}

The solution corresponds to a generalization of the topological star, derived in \cite{Bah:2020ogh,Bah:2020pdz} where the magnetic charges can be different.  The topological star is a solution of five-dimensional Einstein-Maxwell theory that is a consistent truncation of five-dimensional STU supergravity where all gauge fields are equal,  the scalars are trivial, and the Chern-Simons term vanishes.  We retrieve the solution of \cite{Bah:2020ogh,Bah:2020pdz} by considering equal charges, $\delta_1=\delta_2=\delta_3= \delta$,  reparametrizing $(r_\text{B},r_\text{S})=2m (\cosh^2 \delta,\sinh^2 \delta)$ and shifting $r \to r-r_\text{S}$.

The solution ends at $r=r_+=2m$ as the degeneracy of the $\psi$ circle.  The local topology corresponds to an origin in $\IR^2/\mathbb{Z}_k$ if the periodicity of the $\psi$ coordinate,  $\psi = \psi +2 \pi R_\psi$,  satisfies\textsuperscript{\ref{ft1}}
\begin{equation}
k R_\psi \= 4 m \cosh \delta_1 \cosh \delta_2 \cosh \delta_3\,.
\end{equation}

The extremal and BPS limit of the solution is obtained by taking $m\to 0$,  $\delta_I \to \infty$ with $m\sinh 2\delta_I$ kept fixed.  This leads to the singular three-charge BPS black hole.

\subsubsection{(S$_2$): a new soliton}

We apply the $\cP$-group transformations with $\delta_3=\delta_4=0$. The transformed solution is given by\footnote{We have changed $\delta_I \to -\delta_I$ to ensure positive magnetic charges for positive $\delta_I$.}
\begin{align}
ds_5^2 &\=  \frac{\Delta}{(H_1 H_2 V_3)^\frac{2}{3}} \left(d\psi+P_0\cos \theta\, d\phi \right)^2+\frac{(H_1 H_2 V_3)^\frac{1}{3}}{\sqrt{\Delta}}\left[-V_3\sqrt{\Delta}\,dt^2 + \frac{1}{V_3\sqrt{\Delta}}\, ds_3^2 \right],\nn\\
X^I &\= \frac{H_I}{(H_1 H_2 V_3)^\frac{2}{3}}\,,\qquad A^I \=  |\epsilon_{IJ}|  \frac{2n(r-m)\sinh \delta_I  \,\cosh \delta_J}{(r^2-n^2) \,H_J} \, \left(d\psi+P_0\cos \theta\,  d\phi \right) -P_I \cos\theta \,d\phi\,, \nn \\
X^3 & \= \left(H_1 H_2 V_3^4 \right)^\frac{1}{3}\,,\qquad A^3 \= - \frac{Q_3(r-m) \,V_3^2}{r^2-n^2}\,dt\,,
\label{eq:PStar}
\end{align}
where $ds_3^2$ is the invariant base metric \eqref{eq:3dBase}, and we have restricted to $I,J=1,2$.  Moreover, we have defined the fields
\begin{align}
\Delta & \= 1-\frac{2(mr-n^2)}{r^2-n^2}\,,\quad H_I \= 1+\frac{2(mr-n^2)\sinh^2\delta_I}{r^2-n^2}\,,\quad V_3 \= \left(H_1 H_2 - \frac{(r-m)^2 Q_3^2}{(r^2-n^2)^2} \right)^{-\frac{1}{2}}\,,\nn
\end{align}
along with the electric and magnetic conserved charges:
\begin{equation}
P_0 \= 2n \cosh \delta_1 \cosh \delta_2\,,\qquad P_I \= 2m \cosh \delta_I \sinh \delta_I \,,\qquad Q_3 \= -2 n \sinh \delta_1 \sinh \delta_2\,,
\end{equation}
While all other charges are zero.  When embedded in M-theory \eqref{eq:MTheoryEmbed}, $P_0$ corresponds to the KKm charge, $P_1$ and $P_2$ to M5-brane charges, and $Q_3$ to M2-brane charges, where the M2-branes reside at the intersection of the M5-branes.

The solution satisfies the same charge condition as the $\cW$-solitons and the BPS Gibbons-Hawking center \eqref{eq:ChargesCond} in the extremal limit $|n|=m$. However, for generic $m\geq |n|$, the charges are independent with $|Q_3|<P_0$.  We defer the discussion of the extremal limit of the solution ($|n|=m$), as they will emerge as a special limit of the next solution.  

The four-dimensional metric upon reduction along the $\psi$ circle is already given in \eqref{eq:PStar} in the bracketed term. Thus, the ADM mass reads:
\begin{equation}
M_\text{ADM} \= \frac{m}{2G_4}\left(1+\sinh^2\delta_1 + \sinh^2 \delta_2\right).
\end{equation}

The spacetime remains regular for $r>r_+=m+\sqrt{m^2-n^2}$ since $\Delta$, $H_I$, and $V_3$ are all positive. The regularity of the S$^2$ at $\theta=0,\pi$ imposes the same conditions on periodicity and KKm charge as those for the $\cW$-soliton in \eqref{eq:KKmQuantization}.

Furthermore, the solution terminates at $r=r_+$ ($\Delta=0$) as the degeneracy of the $\psi$ circle. The local topology corresponds to an origin in $\IR^2/\mathbb{Z}_k$ if the periodicity of the $\psi$ coordinate, $\psi = \psi +2 \pi R_\psi$, satisfies\textsuperscript{\ref{ft1}}
\begin{equation}
k R_\psi \= 2(m+\sqrt{m^2-n^2}) \cosh \delta_1 \cosh \delta_2 \,.
\end{equation}
This condition matches exactly with that of the $\cW$-soliton \eqref{eq:RegCondQuantized} when expressed in terms of the quantized KKm charge $N_0$ \eqref{eq:KKmQuantization}, so the same analysis applies.\\

Finally, demonstrating that the present $\cP$-soliton exists within the same range of mass and charges as a non-extremal black hole of STU supergravity is slightly more involved than for the $\cW$-soliton. This complexity arises because no simple limit allows all charges to be equal in magnitude.  However, one can restrict to the simpler case where $P_1=P_2$, implying $\delta_1=\delta_2$.

To investigate further, we considered the general eight-charge non-extremal black hole solution from \cite{Chow:2014cca} and identified a parameter limit where only the four charges $(P_0,P_1,P_2,Q_3)$ remain nonzero and have the correct signs.\footnote{This solution is obtained from \cite{Chow:2014cca} by setting $\delta_1 = \delta_2 = \delta$, $\gamma_3 = \gamma/2$, $\gamma_4=-\gamma'/2$, and $\delta_3=\delta_4=\gamma_1= \gamma_2 = n=0$. Note that the four-dimensional solution in \cite{Chow:2014cca} is given in an electromagnetic dual form compared to ours (see their section 2.5.1). Specifically, their $P_1, P_2$, and $P_3$ correspond to our $Q_1, Q_2$, and $Q_3$,  and vice versa.} While the solution does not yield particularly useful simplifications, the expressions for the mass and nonzero charges are given in terms of the parameters: 
\begin{align}
M_\text{ADM}^\text{BH} &= \frac{m\left(2+(\cosh \gamma+\cosh \gamma')\cosh 2\delta \right)}{4G_4} \,,\quad P^\text{BH}_0 \= m\left( \cosh^2 \delta \sinh \gamma' +\sinh^2 \delta \sinh \gamma \right),\nn \\
P_1^\text{BH} &= P_2^\text{BH} = m (\cosh \gamma+\cosh \gamma') \cosh \delta \sinh \delta , \quad Q_3^\text{BH} \= - m\left( \cosh^2 \delta \sinh \gamma +\sinh^2 \delta \sinh \gamma' \right).\nn
\end{align}
 Clearly, the phase space of this black hole is larger than that of the $\cP$-soliton since there is an additional free parameter, and it is possible to have $|Q^\text{BH}_3| \geq P^\text{BH}_0$. However, we seek to determine whether there exists a regime where the $\cP$-soliton and the black hole coexist. In principle, this requires inverting the parameterization to express the mass as a function of the charges, which unfortunately cannot be done analytically. Nevertheless, a numerical scan of the allowed mass and charge values for the $\cP$-soliton suggests that a black hole always exists with the same mass and charges when the soliton is not too close to extremality ($m>|n|$),  suggesting that these smooth, horizonless geometries are microstates of the corresponding black holes.

\subsubsection{(S$_3$): the extremal limit}
\label{sec:PStarExtr}

The extremal $\cP$-solitons are obtained by applying the $\cP$-group transformation with
$\tanh\delta_4=\tanh\delta_1\tanh\delta_2\tanh\delta_3$ and $|n|= m$. The transformed solution can be recast into a metric, scalars, and gauge fields as in \eqref{eq:WStar}, where the fields are expressed in terms of four arbitrary charges $(P_0,P_1,P_2,P_3)$ as follows:\footnote{We have reparametrized from $\delta_{I}$ to the magnetic charges, $P_I= -m \sinh2\delta_I, \quad P_0=\pm m\sqrt{1+\cosh2\delta_1\cosh2\delta_2+\cosh2\delta_2\cosh2\delta_3+\cosh2\delta_3\cosh2\delta_1}$, and shifted the radial coordinate $r \to r+m$ so that the coordinate singularity is at $r=0$.}
\begin{equation}
\begin{aligned}
\Delta &\=\frac{r}{r+4M},\qquad H_I \= 1-\frac{4M P_I^2}{(P_0^2+P_I^2)\,(r+4M)}, \\ 
  \chi &\=\frac{ Q_0}{r +4M},\qquad A_t^I \= -\frac{Q_I}{r+4M}, \qquad  A_\psi^I \=\frac{4M P_0 P_I}{r(P_0^2+P_I^2)+4MP_0^2}. \label{eq:FieldPsolEx}
\end{aligned}
\end{equation}
The electric charges, $Q_0$ and $Q_I$, are determined by the same relation as the $\cW$-soliton \eqref{eq:ChargesCond}, and the parameter $M$ leading to the ADM mass:
\begin{equation}
M \= G_4 M_\text{ADM} \= \frac{1}{4} \sqrt{P_0^2+P_1^2+P_2^2+P_3^2+Q_0^2+Q_1^2+Q_2^2+Q_3^2}.
\end{equation}
The extremal bound in terms of conserved charges is reminiscent of the extremal bound for a static, charged black hole, differing only by an overall factor. For instance, taking all electric and magnetic charges equal, one obtains the extremal bound $G_4 M_\text{ADM} = \frac{1}{2} \sqrt{P^2 + Q^2}$, which corresponds to the extremal bound of the dyonic Reissner-Nordström black hole, but with an additional factor of $\tfrac{1}{2}$.\\

The regularity of the S$^2$ at $\theta=0$ and $\pi$ for all radii requires the same periodicity lattice as in \eqref{eq:Periodicity} and the quantization of the KKm charge $P_0 = \frac{1}{2} N_0 R_\psi$, where $N_0$ is an integer.

Unlike the non-extremal solution, where the spacetime terminates as an origin in $\IR^2$ where the $\psi$ circle shrinks, the extremal solution results in the entire sphere and $\psi$ direction shrinking to form an origin in $\IR^4$. The coordinate singularity is at $r=0$ and the metric along the $(\psi,r,\theta,\phi)$ space takes the form:
\begin{equation}
ds_{\psi,r,\theta,\phi}^2 \,\propto\, \frac{r}{P_0^2}\left(d\psi+P_0 \cos\theta d\phi\right)^2+\left[\frac{dr^2}{r}+r\,(d\theta^2+\sin\theta^2 d\phi^2\,)\right].
\end{equation}
This metric describes an $\IR^4$, with coordinates $\rho=r^2/4$, $\tau=2\theta$, and $\varphi_\pm = \frac{1}{2} \left(\frac{\psi}{P_0} \pm \phi \right)$. Due to the angular identifications, the $r=0$ locus corresponds to $\mathbb{R}^4/\mathbb{Z}_{N_0}$, characteristic of the regular origin of a Taub-NUT space.\\

BPS solutions of five-dimensional supergravity admitting a timelike Killing vector have been classified in \cite{Gauntlett:2002nw,Gutowski:2004yv,Bena:2004de,Bena:2005va}. These solutions are determined by eight harmonic functions $(Z_0,K^1,K^2,K^3,L_1,L_2,L_3,M)$ defined on a flat three-dimensional base. At first glance, our extremal solution appears to involve non-harmonic fields, as shown in \eqref{eq:FieldPsolEx}, and thus seems incompatible with the BPS ansatz of \cite{Gauntlett:2002nw,Gutowski:2004yv,Bena:2004de,Bena:2005va}. However, this ansatz assumes a time direction such that $\cV=\partial_t$ where $\cV$ is the timelike Killing vector constructed from the single Killing spinor. It is possible that our solution corresponds instead to a shifted Killing vector of the form  $\cV = \partial_t - \frac{1}{c}\, \partial_\psi$ where $c$ is a constant.  In that case, the solution can be brought into the BPS form by expressing it as a $(t,\psi+c t)$ fibration over a flat three-dimensional base.

Following this approach,  we found that the extremal $\cP$-soliton fits within the BPS ansatz of \cite{Gauntlett:2002nw,Gutowski:2004yv,Bena:2004de,Bena:2005va} and corresponds to a BPS solution of five-dimensional supergravity once the metric and gauge fields \eqref{eq:WStar} are rewritten as:
\begin{align}
ds_5^2 &= \frac{-1}{(Z_1 Z_2 Z_3)^\frac{2}{3}} \left(dt+\mu \(d\psi +c\, dt+ \omega_\psi\)  \right)^2 +(Z_1 Z_2 Z_3)^\frac{1}{3}\left[ \frac{1}{Z_0} \,\(d\psi +c\, dt+ \omega_\psi\)^2 +Z_0 \,ds(\IR^3)^2\right], \nn \\
A^I &= \frac{1}{Z_I}  \left(dt+\mu \(d\psi +c\, dt+ \omega_\psi\) \right)+\frac{K^I}{Z_0} \(d\psi +c\, dt+ \omega_\psi \)  -P_I \cos \theta d\phi\,, \qquad X^I \= \frac{(Z_1 Z_2 Z_3)^\frac{1}{3}}{Z_I}, 
\end{align}
with the fields expressed in terms of eight harmonic functions as:\footnote{Compared to \cite{Bena:2004de,Bena:2005va,Bena:2007kg}, our conventions have a sign difference.  One can switch between conventions by setting $(\psi,\phi)\to -(\psi,\phi)$, resulting in $(\mu,A_\psi^I,a^I)\to -(\mu,A_\psi^I,a^I)$.}
\begin{equation}
\begin{split}
Z_I=L_I+\frac{|\epsilon_{IJK}|}{2}~ \frac{K^JK^K}{Z_0}, \quad \mu=\frac{M}{2}-\frac{K^IL_I}{2Z_0}-\frac{K^1K^2K^3}{Z_0^2},\quad \omega_\psi = P_0 \cos \theta \,d\phi,
\end{split}
\end{equation}
with
\begin{equation}
c \= \frac{Q_0-(P_1+P_2+P_3)}{\sqrt{P_0^2+P_1^2+P_2^2+P_3^2+Q_0^2+Q_1^2+Q_2^2+Q_3^2}}\,,
\end{equation}
and harmonic functions:
\begin{equation}
\begin{split}
Z_0 &\=\frac{ (P_0 +Q_1+Q_2+Q_3)\, c }{Q_0-(P_1+P_2+P_3)}+\frac{P_0}{r},\qquad K^I \= - c + \frac{P_I}{r}, \\
 M &\= -c +\frac{Q_0}{r},\qquad L_I \= \frac{ (P_0 +Q_1+Q_2+Q_3)\, c }{Q_0-(P_1+P_2+P_3)}+\frac{Q_I}{r}.
\end{split}
\end{equation}
This describes a smooth, horizonless BPS geometry sourced by a single center carrying four electric and four magnetic charges, subject to the constraint \eqref{eq:ChargesCond}. This center is commonly referred to as a \emph{Gibbons-Hawking center}. and such solutions form the basic building blocks of multicenter bubbling geometries that realize BPS black hole microstates in supergravity \cite{Bena:2007kg,Heidmann:2017cxt,Bena:2017fvm,Warner:2019jll}.

The time shift in the Gibbons–Hawking fiber, $\bar{\psi} \equiv \psi + c\, t$,  corresponds to a spectral flow along the $\psi$ direction \cite{Bena:2008wt}.  In the BPS frame $(\bar{\psi},t,r,\theta,\phi)$,  the solution is not asymptotic to a S$^1$ fibration over a four-dimensional Minkowski spacetime due to the scalar $\mu$ approaching a finite constant at large distances. As a result, the BPS frame is not asymptotically static. Returning to the asymptotically static frame requires transforming the metric and gauge fields back to the form in \eqref{eq:WStar} with \eqref{eq:FieldPsolEx}, which makes the BPS structure of the solution less manifest.

The BPS and asymptotically static frames coincide only when $c=0$ which restricts the charges to satisfy:
\begin{equation}
P_0 \=-\frac{\sqrt{P_1}\sqrt{P_2}\sqrt{P_3}}{\sqrt{P_1+P_2+P_3}},\qquad P_1 + P_2 +P_3 \,>\,0\,.
\label{eq:ChargesBPS}
\end{equation}

Moreover,  the rest-mass (or BPS mass),  $M_0$,  can be computed by reducing along $\bar{\psi}$ rather than $\psi$,  yielding
\begin{equation}
 M_0 =\frac{1}{4G_4} (P_0+Q_1+Q_2+Q_3).
\end{equation}
This mass formula is characteristic of intersecting M2 branes with a Kaluza–Klein monopole (KKm) charge, preserving 16 supercharges in M-theory compactified on S$^1\times$T$^6$.

In conclusion, extremal $\cP$-solitons can be interpreted as a well-defined coordinate transformation of non-static BPS solutions in five-dimensional supergravity, sourced by a single Gibbons–Hawking center. This transformation induces a nontrivial time shift in the Gibbons–Hawking fiber, rendering the solution asymptotically at rest.

\section{Discussion and future directions}

This paper provides a new roadmap for generating microstate solutions of non-extremal black holes in supergravity. We have adopted a sigma-model approach that enables the construction of exact solutions in classical theories of gravity and used it to find smooth horizonless geometries in supergravity. The key innovation of our technique is that it operates independently of supersymmetry and BPS equations, making it directly applicable to the non-extremal regime of mass, charges, and spin. At the same time, it preserves the essential ingredients needed to construct vast phase spaces of coherent black hole microstates, as theorized in supersymmetric setups \cite{Gibbons:2013tqa}: topology along extra compact dimensions and electromagnetic fluxes.

We applied this strategy to $\cN=2$ five-dimensional supergravity coupled to two vector multiplets, revisited the sigma model governing stationary solutions with a U(1) isometry and its associated solution-generating techniques, and provided the first Ernst formulation of the model. We have then demonstrated the effectiveness of our approach by constructing new families of non-extremal eight-charge microstate geometries. These solutions are connected to the fundamental building block of BPS multicenter microstate geometries,  the Gibbons-Hawking center,  but significantly depart from it beyond the supersymmetric regime.

Our analysis paves the way for numerous promising research directions. In this paper, we have only begun to explore the potential of the solution-generating techniques for constructing novel physical solutions in supergravity:
\begin{itemize}
\item[•] The static, smooth, horizonless solutions derived in this paper were obtained by applying transformations from four-parametric subgroups of SO(4,4). As discussed in Section \ref{sec:SO44Transfo}, transformations that preserve asymptotic four-dimensional flatness define a larger, twelve-parametric subgroup. Constructing the most general solution by applying transformations from this extended group, parameterized by twelve independent parameters, would be an interesting direction to explore. Furthermore, using normalized Ernst transformations instead of SO(4,4) transformations should be equivalent. However, since the parametrizations of these transformations differ, the resulting solutions could take a significantly simpler form.
\item[•] A crucial next step is the construction of microstate solutions for non-extremal \emph{rotating} black holes, which hold significant phenomenological importance. This can be achieved by applying SO(4,4) or Ernst transformations to a Kerr Taub-bolt seed solution (the Euclidean Kerr-NUT geometry). To date, most of the constructed rotating smooth solutions exhibit over-rotating characteristics, placing them outside the black hole regime of mass, charges, and spin \cite{Jejjala:2005yu,Giusto:2007tt,Katsimpouri:2014ara,Banerjee:2014hza}. The ability to construct exactly static solutions using our approach suggests that we may be able to provide the first large families of rotating smooth geometries that exist in the black hole regime.
\item[•] Another key direction is the transition from microstates sourced by a single bolt to multi-bubble geometries. This can be achieved by applying the inverse scattering method reviewed in this paper or SO(4,4) or Ernst transformations to smooth vacuum multi-bubble solutions such as those derived in \cite{Bah:2021owp}.  The ultimate goal of this approach is to construct neutral, rotating, smooth horizonless geometries supported by electromagnetic fluxes with zero net charges, making them directly comparable to Kerr black holes — similar in spirit to what has been done for Schwarzschild \cite{Bah:2022yji,Bah:2023ows,Dulac:2024cso,Heidmann:2023kry}.
\item[•] In this paper, we focused on solutions asymptotic to four-dimensional Minkowski space, but the same techniques can be applied to different asymptotics, such as five-dimensional flat space, AdS$_2\times$S$^3$, and AdS$_3\times$S$^3$ (when embedded in the type IIB dual frame of Section \ref{sec:typeIIB}), enabling a holographic description of these new non-supersymmetric states. Working with new asymptotics is challenging when using SO(4,4) transformations, as it requires identifying the SO(4,4) matrices that preserve the asymptotic coset matrix $\cM$ (see Section \ref{sec:SO44Transfo}). However, it is significantly simpler using normalized Ernst transformations,  \eqref{eq:NormalizedScale} and \eqref{eq:NormalizedErnst},  demonstrating the effectiveness of the Ernst formulation of the sigma model. 
\item[•] In parallel, an important question is the effect of these transformations on the large families of BPS microstates already constructed in the literature. Many BPS microstate geometries are stationary and depend only on three variables \cite{Bena:2007kg,Heidmann:2017cxt,Bena:2017fvm,Warner:2019jll} (including some classes of superstrata \cite{Heidmann:2019xrd}). Applying SO(4,4) or Ernst transformations to these solutions should preserve their inner topology, meaning they remain extremal. However, it could still make them non-BPS, providing a simple way to generate large families of non-supersymmetric microstates.
\end{itemize} 

To gain deeper insight into the non-BPS geometries that can be constructed through this program, several key questions regarding their dynamics and gravitational signatures must be addressed:
\begin{itemize}
\item[•] The first major question concerns stability. This includes investigating classical instabilities, as done for similar smooth horizonless solutions \cite{Cardoso:2005gj,Bena:2024hoh,Dima:2024cok,Bianchi:2024vmi,Dima:2025zot}, and thermodynamic stability \cite{Bah:2021irr}. If these solutions prove to be unstable, quantizing weakly coupled quantum fields in these microstate backgrounds could yield profound insights into Hawking radiation and the black hole information paradox. A natural approach would be to first solve the free field equations in these backgrounds, determine the semiclassical potential, and ultimately compute the transition amplitude using the WKB approximation, as was done to study the instability of other non-BPS smooth geometries in string theory \cite{Chowdhury:2007jx}.
\item[•] The second question concerns gravitational signatures and how these geometries compare to black holes.  As a non-exhaustive list, this includes deriving gravitational multipole moments \cite{Bena:2020uup}, tidal Love numbers \cite{Consoli:2022eey}, quasi-normal modes and responses to perturbations \cite{Heidmann:2023ojf,Dima:2025zot}, probe geodesics and imaging simulations \cite{Bacchini:2021fig,Heidmann:2022ehn}, and tidal forces \cite{Tyukov:2017uig,Martinec:2020cml,Heidmann-Patashuri}. Comparing these properties between non-BPS microstate geometries and their corresponding non-extremal black holes could provide valuable insights into the nature of these exact, yet likely atypical, microstructures of non-extremal black holes.
\end{itemize}

\section*{Acknowledgements}

We are grateful to Ibrahima Bah, Iosif Bena,  Gérard Clément, Raphaël Dulac,  Bogdan Ganchev,  Samir Mathur,  Paolo Pani, Jorge Santos and Helvi Witek for useful discussions.  The work of SC and PH is supported by the Department of Physics at The Ohio State University.

\bibliographystyle{utphys}      

\bibliography{microstates}       


\end{document}